\documentclass[a4paper,11pt]{article}
\pdfoutput=1 

\usepackage{jheppub} 

\usepackage{orcidlink} 
\usepackage[T1]{fontenc} 
\usepackage{amsmath}
\usepackage{amsfonts}
\usepackage{amssymb}
\usepackage{bm}
\usepackage{bbm}
\usepackage{physics}
\usepackage{braket}
\usepackage{array}
\usepackage{upgreek}
\usepackage{supertabular}
\usepackage{graphicx}
\usepackage{caption}
\usepackage{lipsum}
\usepackage{hhline}
\usepackage{afterpage}
\usepackage{hvfloat}
\usepackage{mathtools}
\usepackage{mhchem}
\usepackage[compat=1.1.0]{tikz-feynman}
\usepackage[capitalize]{cleveref}
\usepackage{setspace}
\usepackage{etoolbox}

\newcommand{\qft}{\text{QFT}}
\newcommand{\itf}{\text{ITF}}

\newcommand{\greenn}[1]{\left(\frac{1}{{\omega^2_{n_{#1}}}+\varepsilon_{#1}^2}\right)}

\newcommand{\greenna}[2]{\left(\frac{#1}{{\omega^2_{n_{#2}}}+\varepsilon_{#2}^2}\right)}

\newcommand{\DD}[2]{\frac{\dd[{#1}]{#2}}{(2 \pi)^{#1}}}

\newcommand\arraybslash{\let\\\@arraycr}

\newcommand{\rge}
{
	\left[  \frac{\partial}{\partial \ln \mu}+ \beta(g) \frac{\partial}{\partial g}-n \gamma(g) + \gamma_m(g) \frac{\partial}{\partial \ln m}  \right]
}
\newcommand{\TheVacuuma}
{
	\begin{tikzpicture}
		\draw(-6,-11.75) circle(0.25);
	\end{tikzpicture}
}
\newcommand{\TheVacuumb}
{
	\begin{tikzpicture}
		\draw(-6,-11.75) circle(0.25);
		\draw(-5.5,-11.75) circle(0.25);
	\end{tikzpicture}
}
\newcommand{\TheVacuumCounterterm}
{
	\begin{tikzpicture}
		\draw(-6,-11.75) circle(0.25);
		\draw (-6,-12) node[cross,rotate=0] {};
	\end{tikzpicture}
}
\newcommand{\TheTadpole}
{
	\begin{tikzpicture}
		\draw(-6,-11.75) circle(0.25);
		\draw(-6.5,-12) -- (-5.5,-12);
	\end{tikzpicture}
}

\newcommand{\TheScatter}
{
	\begin{tikzpicture}
		\draw(-6,-11.75) circle(0.25);
		\draw(-6.25,-11.75) -- (-6.43,-11.6);
		\draw(-6.25,-11.75) -- (-6.43,-11.92);
		\draw(-5.75,-11.75) -- (-5.57,-11.6);
		\draw(-5.75,-11.75) -- (-5.57,-11.92);
	\end{tikzpicture}
}
\newcommand{\TheDoublebubble}
{
	\begin{tikzpicture}
		\draw(-6,-11.75) circle(0.25);
		\draw(-6.5,-12) -- (-5.5,-12);
		\draw(-6,-11.25) circle(0.25);
	\end{tikzpicture}
}

\newcommand{\TheSunrise}
{
	\begin{tikzpicture}
		\draw(-6,-12) circle(0.25);
		\draw(-6.5,-12) -- (-5.5,-12);
	\end{tikzpicture}
}

\newcommand{\Twopointsimple}
{
	\begin{tikzpicture}
		\draw(-6.0,-12) -- (-5.5,-12);
	\end{tikzpicture}
}

\newcommand{\countertermone}
{
	\begin{tikzpicture}
		\draw[black,fill=black] (-6,-12) circle(0.5ex);
		\draw (-6,-12) node[cross,rotate=0] {};
	\end{tikzpicture}
}
\newcommand{\countertermbtwo}
{
	\begin{tikzpicture}
		\draw(-6,-11.75) circle(0.25);
		\draw(-6.5,-12) -- (-5.5,-12);
		\draw[black,fill=black] (-6,-11.5) circle(0.5ex);
	\end{tikzpicture}
}
\newcommand{\countertermtwo}
{
	\begin{tikzpicture}
		\draw(-6,-11.75) circle(0.25);
		\draw(-6.5,-12) -- (-5.5,-12);
		\draw (-6,-11.5) node[cross,rotate=0] {};
	\end{tikzpicture}
}
\newcommand{\countertermthree}
{
	\begin{tikzpicture}
		\draw(-6,-11.75) circle(0.25);
		\draw(-6.5,-12) -- (-5.5,-12);
		\draw[black,fill=black] (-6,-12) circle(0.5ex);
	\end{tikzpicture}
}

\newcommand{\masscounterterm}
{
	\begin{tikzpicture}
		\draw(0.0,0) -- (1.0,0);
		\draw (0.5,0) node[cross,rotate=0] {};
	\end{tikzpicture}
}
\newcommand{\momentumcounterterm}
{
	\begin{tikzpicture}
		\draw(0.0,0) -- (1.0,0);
		\draw (.5,0) circle (3pt);
	\end{tikzpicture}
}

\usepackage[compat=1.1.0]{tikz-feynman}
\usepackage{natbib}
\usepackage{bm}
\usepackage{fdsymbol}
\usetikzlibrary{shapes.misc}
\usepackage[font=small,labelfont=bf,
justification=raggedright,
format=plain]{caption} 
\setcitestyle{author,square}
\tikzset{cross/.style={cross out, draw=black, minimum size=2*(#1-\pgflinewidth), inner sep=0pt, outer sep=0pt},
	cross/.default={5pt}}

\title{\boldmath Self-Energy Approximation for the Running Coupling Constant in Thermal $\phi^4$ Theory using Imaginary Time Formalism}

\author[a,b]{K. Arjun\orcidlink{0000-0001-9260-7050},}
\author[b]{A M Vinodkumar\orcidlink{0000-0002-8204-7800},}
\author[b]{Vishnu Mayya Bannur}
\author[c] {and Munshi G. Mustafa\orcidlink{0000-0002-7874-6932} \footnote{Ex-senior Professor, Theory Division, Saha Institute of Nuclear Physics, Homi Bhabha National Institute, 1/AF, Bidhannagar, Kolkata 700064.}}

\affiliation[a]{Srinivasa Ramanujan Institute For Basic Sciences, Kerala, India - 686501.}
\affiliation[b]{Department of Physics, University of Calicut, Kerala, India - 673635.}
\affiliation[c]{Department of Physics, IIT Bombay, Pawai, Mumbai 400076,
	India.}
\emailAdd{arjunk\_dop@uoc.ac.in}

\abstract{The running coupling constant is calculated using the imaginary time formalism (ITF) of thermal field theory under the self-energy approximation. In the process, each Feynman diagram in thermal field theory is rewritten as the summation of non-thermal diagrams with coefficients that are functions of mass and temperature. By employing the same mass scale and coupling constant for both the non-thermal QFT and ITF, we derive a relation between them. Also,  we calculate the self-energy using ITF, which is equated to the same as that of non-thermal QFT under the zero external momentum limit. This can provide a new expression for the coupling constant. Combining this result with the $\beta(g)$ and $\gamma_m(g)$ function relations of the renormalization group equations gives rise to a thermal-dependent coupling constant and running mass. Using these results, the free energy density is evaluated for two-loop order and compared with quasiparticle model.
}

\begin{document} 
\maketitle
\flushbottom
\section{Introduction}
The running coupling constants in various field theories play a pivotal role in describing the strength of interactions and developing equations of state \cite{Arjun2022,Kapusta2006,Khanna2009,Millington2014,Sarkar2010,Padmanabhan2016}. The application of such coupling constants in various phenomenological models has produced results that align with lattice data describing the interactions of particles under extreme conditions, resembling some aspects of the primordial universe \cite{Yagi2005a,Bannur2007,Bannur2007a,Bannur2007b}. \\

In 1949, Dyson established the systematic approach to multiplicative renormalization, a method through which infinities are eliminated within the context of QED \cite{Dyson1949}. Further, concepts such as the renormalization group equation were introduced by Stueckelberg and Petermann \cite{Stueckelberg1953}, and Gell-Mann and Low \cite{GellMann1954}. A rigorous mathematical procedure for higher-order calculations using the method of counterterms, which involves a structured process for subtracting divergences, were formulated by Bogoliubow and Parasiuk \cite{Bogoliubow1957} and Hepp \cite{Hepp1966}. Later, the minimal subtraction scheme (MS Scheme) \cite{Hooft1973,Hooft1972,Weinberg1973} was developed. This scheme is an important renormalization technique in field theory, wherein Feynman integrals are computed using dimensional regularization. This approach offers several advantages, including its simplicity in identifying diverging terms and its ability to regularize nonabelian field theories while preserving gauge symmetries, among others \cite{Hooft1973,Hooft1972,Bollini1972}. Callan \cite{Callan1970} and Symanzik \cite{Symanzik1970} introduced a differential equation to study the behavior of the vertex functions under various scale approximations in the context of cutoff regularization. Subsequently, the formalism was extended to encompass the MS Scheme as well \cite{Kleinert2001,Padmanabhan2016}. \\

In the 1950s, Matsubara \cite{Matsubara1955} developed a systematic approach to describe and evaluate ensemble expectation values using quantum field theory at finite temperature, which later became popularly known as the imaginary time formalism. This method employs Wick rotation and treats the time component as an imaginary value. While there are some differences between the imaginary time formalism and zero-temperature quantum field theory, one could see a one-to-one correspondence between them, especially in the cases of Green's functions and diagrammatic representations \cite{Kapusta1979,Khanna2009,Millington2014}. 
However, the requirement of a theory for nonequilibrium systems has led to the foundation of real-time formalism at finite temperature \cite{Keldysh1964,Schwinger1961,Mahanthappa1962,Bakshi1963}  using the closed-time path formulation. One of the important features of real-time formalism (RTF) is the doubling of the degrees of freedom. Therefore, the Green functions are expressed in $2 \times 2$ matrices. Some interesting works in RTF can be found in \cite{CaronHuot2009,Santos2020,Lundberg2021,Carrington1999,Bergerhoff1998,Aurenche1992,Landsman1987,Ezawa1991,Bellac2000,Khanna2009,Millington2014}. The investigation of the $\phi^4$ coupling constant and related parameters can be explored in both real time and imaginary time. However, in this theoretical work, we prioritize the ITF. A recent and comprehensive analysis of the imaginary time formalism can be found in \cite{Mustafa2023}, which details its theoretical foundations and applications \\

References \cite{Kapusta2006,Matsumoto1984,Fujimoto1986,Fujimoto1988} discuss Renormalization Group Equations (RGE) within the framework of finite temperature field theory, in which \cite{Matsumoto1984,Fujimoto1986,Fujimoto1988} introduce multiple sets of RGE equations to calculate temperature-dependent coupling constant and mass. This is achieved by adding an additional parameter to the Lagrangian, alongside the mass scale, which results in the development of a new set of renormalization group equations. The application of both sets of RGEs to the vertex function establishes new relations among various parameters. In reference \cite{Fujimoto1986}, following the imposition of specific renormalization conditions, the authors analyzed the coupling constant and effective mass. They noted that as the temperature increases, the coupling constant decreases, while the effective mass increases. These works \cite{Fujimoto1988,Baier1990,Nakkagawa1987,Nakkagawa1988} collectively suggest that the nature of the coupling constant heavily depends on the choice of the vertex function.\\ 

In the 1990s, Braaten and Pisarski \cite{Braaten1990} introduced a resummation scheme using the concept of hard-thermal loops. Through its application, they provided a resummation of QCD thermal perturbation theory.

From 1992 onwards, multiple studies and calculations have analyzed the pressure and free energy density for the massless $\phi^4$ theory under various approximations, as outlined in references \cite{Frenkel1992,Arnold1994,Parwani1995a,Andersen2009}. Similarly, Andersen \emph{et al}. \cite{Andersen2000} calculated the free energy for the massive $\phi^4$ theory up to three-loop order. In recent years, modern thermal perturbation theory, particularly the hard thermal loop (HTL) approximation, has garnered new interest for calculating key thermodynamic quantities such as pressure and free energy density in high-temperature quantum field theories. This framework plays a crucial role in understanding plasmas within quantum electrodynamics (QED) and quantum chromodynamics (QCD). A thorough discussion on HTL can be found in \cite{Haque2025}. Some recent interesting studies in both thermal and non-thermal field theory can be found in references \cite{Arjun2022,Romatschke2023,Romatschke2024a,Romatschke2024b,Bandyopadhyay2016,Bandyopadhyay2016a}. \\

Several works utilize the self-energy as an effective mass, particularly in the case of massless Lagrangian approximations under different conditions \cite{Parwani1995a,Peshier1998}. These approximations have led to the emergence of new relations between mass and coupling constants, enabling the evaluation of diverse thermodynamic quantities and providing novel insights in the field.\\

In this study, we explore the non-linear temperature dependence of the mass scale, coupling constant, and running mass. This is significant because phenomenological models often use running coupling constants to introduce quantum effects and determine the effective mass of interactions. To incorporate thermal dependence in the coupling constant, these models typically assume the scale parameter to be a linear function of temperature. However, earlier works, such as \cite{Matsumoto1984,Fujimoto1986}, have explored alternatives to this assumption, particularly within the framework of thermal field theories. For instance, \cite{Matsumoto1984,Fujimoto1986,Fujimoto1988} employed multiple renormalization group equations (RGEs) to calculate the temperature-dependent coupling constant, particularly in the case of \( \phi^4 \) theory, suggesting the possibility of a non-linear temperature dependence on the scale parameter.

We show that without relying on multiple RGEs, the established RGE, as in \cref{onerge}, when simultaneously applied to both thermal and non-thermal vertex functions, can reveal the relationship between the mass scale, running mass, and coupling constant as functions of temperature. While a similar approach is followed in \cite{Arjun2022}, our work differs by exploring the less-studied case where the self-energy computed in ITF and QFT becomes equal. This approximation, referred to as the self-energy approximation (SEA), provides a novel perspective on these relationships. However, it is imperative that this approximation remains consistent with the Renormalization Group Equation (RGE). In this work, we focus on the Lagrangian of the following form:
\begin{equation}
	\mathcal{L}= \frac{1}{2} \left\lbrace \partial_\mu \phi \partial^{\mu} \phi - m^2 \phi^2 \right\rbrace - \frac{\lambda}{4 !} \phi^4
\end{equation} 

It is already known that the renormalization group functions ($\beta(g)$, $\gamma(g)$, $\gamma_m(g)$) associated with the renormalization group equation for the $\phi^4$ theory are identical for both non-thermal quantum field theory and the imaginary time formalism up to two-loop order \cite{Arjun2022,Kapusta2006,Laine2016}.
The renormalization group equation follows a general structure of
\begin{equation}
	\begin{aligned}\label{onerge}
		\rge \ \ \bar{\Gamma}^{(n)} = 0 
	\end{aligned}
\end{equation}
where $g$ is the running coupling constant, $m$ is the running mass, and $\mu$ is the mass scale.
\\
In \cref{onerge}, \(\bar{\Gamma}^{(n)}\) corresponds to the \(n\)-point proper vertex function (PVF) \cite{Kleinert2001}. However, in this work, we are mainly dealing with two- and four-point vertex functions. The two-point vertex function is defined as the zeroth-order Feynman diagram minus the self-energy ($\Sigma$). i.e.,
\begin{equation}
	\Gamma^{(2)} = \left( \Twopointsimple \right)^{-1} - \Sigma = K^2+m^2-\Sigma 
\end{equation}
The finite two-point vertex function \(\bar{\Gamma}^\text{(2)}\) is defined as 
\begin{equation}
	\begin{aligned} \label{thefinite}
		\bar{\Gamma}^{(2)} &= \Gamma^{(2)} - \mathcal{K} \left( \Gamma^{(2)} \right)
	\end{aligned}
\end{equation}
Here, \(\mathcal{K}\) is an operator that extracts the pure pole terms from the dimensionally regularized integral for each diagram. Applying \(\mathcal{K}\) in \cref{thefinite} ensures it is finite. The detailed diagrammatic expansion of the self-energy for various orders can be found in \cref{regularization}.

The usual procedure to find the running coupling constant and running mass is through the beta function ($\beta(g)$) and gamma function ($\gamma_m(g)$) relations. However, in order to determine the temperature-dependent coupling constant, one might need one additional equation that relates the coupling constant and temperature. 

To enhance the clarity of calculations using diagrammatic techniques, we will introduce some convenient notations.
Throughout this work, we will utilize the subscripts QFT and ITF to denote Feynman diagrams and vertex functions expressed in non-thermal \emph{quantum field theory} and finite temperature \emph{imaginary time formalism}, respectively. The QFT subscript denotes the \emph{vacuum part} expression (i.e., $T=0$ field theory in four Euclidean dimensions \cite{Kapusta2006}).  
 
  The zeroth-order Feynman diagrams with subscripts QFT and ITF are defined as
\begin{equation}\label{simpleqft}
	\left( \Twopointsimple \right)^{-1}_\qft = K^2+m^2 
\end{equation}
and
\begin{equation}\label{simpleitf}
	\left( \Twopointsimple \right)^{-1}_\itf =  (2 \pi n_k T)^2+k^2+m^2
\end{equation}
The substitution of \( K = [2 \pi n_k T, k] \), i.e., \( K_0 = 2 \pi n_k T \), makes \cref{simpleqft,simpleitf} equal. 
The use of the subscripts QFT and ITF can be further explained using diagrams $\TheTadpole$ and $\TheScatter$.
If we denote \( p^2 + m^2 = \varepsilon_p^2 \), \( n_B(x) = \left[\exp(x) - 1\right]^{-1} \), \( \beta = 1/T \), and \( \omega_n = 2 \pi n T \), then the diagrams in the ITF and QFT representations can be related as shown below.

The expression for the tadpole diagram in ITF is
\begin{equation}\label{vacuumtad3}
	\begin{aligned}
		\TheTadpole_\itf &= -\lambda \ T \sum_{n_p= -\infty}^\infty \int \DD{3}{p} \greenn{p}  
	\end{aligned}
\end{equation}
and in QFT, it is 
\begin{equation}
	\begin{aligned}\label{vacuumtad2}
		\TheTadpole_\qft &= -\lambda \int \frac{1}{\varepsilon_{P}^2} \DD{4}{P} = -\lambda \int \frac{1}{P^2+m^2} \DD{4}{P} \text{.}
	\end{aligned}
\end{equation}
Both these diagrams can be connected as
\begin{equation}\label{vacuumtad1}
	\TheTadpole_\itf = \TheTadpole_\qft -\lambda  S_{1}(m,T)
\end{equation}
where
\begin{equation}
	\begin{aligned}
		S_1(m,T)=\int \frac{n_B(\beta \varepsilon_p)}{\varepsilon_p} \frac{\dd[3]p}{(2 \pi)^3} \text{  .}
	\end{aligned}
\end{equation}

Similarly, the scattering diagram is defined in ITF as
\begin{equation}
	\TheScatter_\itf  = 	\int \DD{3}{p} \  \sum_{n_p= -\infty}^\infty \frac{\lambda^2 T}{\omega_{n_p}^2+\varepsilon_p^2} \frac{1}{\omega_{n_p-n_r}^2+\varepsilon_{p-r}^2} \text{.}
\end{equation}
In QFT, the corresponding expression for the diagram is
\begin{equation} \label{vacuumscatter2}
	\TheScatter_{\substack{\qft }} =   \int \frac{\lambda^2}{P^2+m^2}\frac{1}{(P-R)^2+m^2} \DD{4}{P} \text{  .}
\end{equation}
Both these diagrams can be connected at $R_0=\omega_{n_r}=2 \pi n_r T$ as
\begin{equation} \label{vacuumscatter1}
	\ \TheScatter_{\itf} = 
	\TheScatter_{\qft,R_0=\omega_{n_r}} + \lambda^2 \ W(r,n_r)
\end{equation}
with
\begin{equation}
	\begin{aligned}
		W(r,n_r) =  		\sum_{\sigma,\sigma_1=\pm 1} \int  \frac{n_B(\beta \varepsilon_p)}{2 \varepsilon_p \varepsilon_{p+r}} \frac{1}{\sigma_1 \varepsilon_p+\varepsilon_{p+r}+ i \sigma \omega_{n_r}} \DD{3}{p} \text{  .}
	\end{aligned}
\end{equation}

The important part is that having the same structure for the renormalization coefficients and renormalization function equations as in non-thermal QFT is not enough to derive the temperature-dependent running coupling constant.  To derive a thermal-dependent coupling constant, \cite{Arjun2022} took a novel approach.
This approach establishes a connection between non-thermal QFT and the ITF, assuming both share the same mass scale (\(\mu\)) and coupling constant (\(g\)). Hereafter, we will refer to this approach as SMC, which stands for Same Mass Scale and Coupling. \\

In this work, we adopt the term \emph{two-point vertex function} to denote $\Gamma^{(2)}$, following the convention in \cite{Kleinert2001}. While in more recent literature, the loosely related term \emph{propagator} is also frequently used, the two terms refer to similar concepts with slight differences in interpretation depending on context. For the purposes of this discussion, we will adhere to the terminology \emph{two-point vertex function} for consistency. 

In \cite{Arjun2022}, within the SMC approximation, each Feynman diagram in ITF was expressed as the summation of Feynman diagrams in QFT, with coefficients that depend on temperature and mass. In this approach, renormalization scale independence through the RGE is imposed on the two and four-point functions for both thermal and non-thermal contributions simultaneously. This results in the usual running of renormalized parameters as a function of the renormalization scale. However, the additional constraint from the thermal part can only be satisfied by incorporating a temperature dependence of \(\mu\). In doing so, \cite{Arjun2022} obtains an additional equation that involves \(\mu, m, g, \text{ and } T\). \\ 
The process described above can be expressed mathematically as follows: the usual RGE equation demands, following the renormalization procedure (\(\lambda = g \mu^\epsilon\)), that in the case of the finite, two-point, two-loop vertex function ($\bar{\Gamma}^{(2)}$), 

\begin{equation}\label{tworge}
	\rge \ \bar{\Gamma}^{(2)}_\itf (m,g,\mu,T)\approx 0  \\
\end{equation}
and
\begin{align}\label{tworge2}
	\rge \ \bar{\Gamma}^{(2)}_\qft (m,g,\mu) \approx 0
\end{align}
for ITF and QFT, respectively \cite{Kapusta2006,Kleinert2001}. Since the renormalized vertex function ($\bar{\Gamma}$)  can be expressed as a combination of diagrams, the SMC approach, along with some Feynman diagram manipulations, allows us to connect the ITF and QFT vertex functions under the zero external momentum limit \cite{Arjun2022} as 
\begin{equation}\label{vertex}
	\bar{\Gamma}^{(2)}_\itf = \bar{\Gamma}^{(2)}_\qft (m,g,\mu)+\bar{\Gamma}^\text{(2)diff } (m,g,\mu,T) \text{ .}
\end{equation}
It should be noted that \(\bar{\Gamma}^{\text{(2)diff}} = \bar{\Gamma}^{\text{(2)diff}}(m, g, \mu, T)\) has explicit dependence on temperature \(T\), mass \(m\), coupling constant \(g\), and mass scale \(\mu\). In contrast, \(\bar{\Gamma}^\text{(2)}_\qft = \bar{\Gamma}^\text{(2)}_\qft(m, g, \mu)\) has explicit dependence only on \(m\), \(g\), and \(\mu\). This does not rule out the possibility that \(\bar{\Gamma}_\qft(m, g, \mu)\) can have temperature dependence non-explicitly via a temperature-dependent mass scale \(\mu\) and other parameters.\\ 

When the RGE equation is applied at the two-loop order for the two-point function and combined with \cref{tworge,tworge2,vertex}, it results in the obvious relation that,
\begin{equation}\label{turn}
	\rge \ \bar{\Gamma}^{\text{(2)diff}} (m,g,\mu,T)\approx 0
\end{equation}
This expression indicates that the mass scale $\mu$ needs to be temperature-dependent in such a way that it satisfies \cref{turn}, along with the running mass and coupling constant.
A detailed explanation can be found in \cite{Arjun2022} and in \cref{regularization}.
\\

\cref{turn} can be analyzed in two ways:
\begin{enumerate}
	\item $\bar{\Gamma}^{\text{(2)diff}} \neq 0$, extensively examined and discussed in \cite{Arjun2022}, and
	\item $\bar{\Gamma}^{\text{(2)diff}}=0$, investigated within this study.
\end{enumerate}

The former case 1) has resulted in a new relation involving the coupling constant $g$, running mass $m$, temperature $T$, and mass scale $\mu$. Combining this with the renormalization group function ($\beta(g), \ \gamma_m(g)$) equations has revealed how the coupling constant, running mass, and mass scale vary with temperature. Later, leveraging these relations and combining them with a quasiparticle model has demonstrated that the pressure reaches the Stefan-Boltzmann limit \((\pi^2/90) T^4\)  at high temperatures \cite{Arjun2022}.\\

In this work, we are exploring another possibility that has been less explored, i.e., the latter case where 
\begin{equation}\label{zerocase}
	\bar{\Gamma}^\text{(2)diff}(m,g,T,\mu) = 0
\end{equation}
, which is consistent with the RGE associated  \cref{tworge,tworge2,vertex,turn}. However, the relation connecting the vertex function and self-energy is \(	\Gamma^{(2)}=\left( \Twopointsimple \right)^{-1}-\Sigma^{(2)}
\), where \( \Sigma \) represents self-energy and $\left( \Twopointsimple \right)^{-1}=K^2+m^2$. Therefore, in the zero external momentum limit, using the relation between the two-point vertex function and the self-energy, it is possible to write \cref{vertex} as follows
\begin{equation}\label{vertex2}
	\Sigma_\itf = \Sigma_\qft - \bar{\Gamma}^{\text{(2)diff}}
\end{equation}

This, in turn, from \cref{vertex2}, indicates that when \( \bar{\Gamma}^\text{(2)diff} = 0 \), the self-energy in ITF and QFT becomes equal (see \cref{regularization} for detailed derivation).  From now onwards, we refer to this approximation as the Self-Energy Approximation (SEA). To simplify calculations, we perform them at the limit of zero external momentum. In SEA, the self-energy has a non-explicit dependence on temperature, such that the values of \( \mu \) and other parameters are constrained in a way that \( \bar{\Gamma}^\text{(2)diff}(m, g, T, \mu) = 0 \) always. When \cref{zerocase} is combined with the $\beta(g)$ and $\gamma_m(g)$ relations and solved simultaneously, it results in a temperature-dependent running coupling constant $g$, running mass $m$, and mass scale $\mu$. \\

Throughout this paper, we utilized the Euclidean momentum representation with \(K = [\omega_n, k]\) to express \(K^2 = \omega_n^2 + k^2\). Some of the Feynman diagram integral results and formulas derived and utilized in this work may be found in other publications \cite{Arjun2022,Kapusta2006, Bugrij1995, Andersen2000, Andersen2001a, Kleinert2001}. The mathematical conventions utilized in this work are established from references \cite{Arjun2022,Kleinert2001}.\\

This work is organized as follows: In \cref{regularization}, we establish a connection by writing down all essential diagrams and counter diagrams in the finite temperature imaginary time formalism (ITF) in terms of the corresponding diagrams in quantum field theory (QFT).  In \cref{selfenergy}, a new relation for the coupling constant, mass scale, running mass, and temperature is derived using renormalization group function relations and self-energy approximation (SEA). The evaluation of the running mass, coupling constant, and mass scale is then combined with the renormalization group functions.

Over the past few decades, the study of Quark-Gluon Plasma (QGP) has advanced significantly through various phenomenological models, including quasiparticle models \cite{Bannur2007,Bannur2007a,Bannur2007b,Peshier1994,Peshier1996,Peshier1998} that account for the non-ideal behavior of QGP observed in both lattice QCD simulations and heavy-ion collision experiments. These models often approximate the effective mass of the interacting particles as a function of the coupling constant, typically expressed as \( m(T) \propto g(T) T \), where \( g(T) \) is the temperature-dependent coupling constant. While this paper primarily focuses on the thermodynamic properties of $\phi^4$ theory, it also draws upon insights from these quasiparticle models to establish a comparative framework. Specifically, we examine the free energy density of the massive $\phi^4$ theory and its relation to the quasiparticle model. In \cref{results}, the values of the running mass, coupling constant, and mass scale are applied to these models, and the pressure and free energy are evaluated for various orders and parameters.
\section{Regularization}\label{regularization}
The diagrams in $\phi^4$ theory can be described as compositions of fundamental, non-trivial Feynman diagrams known as one-particle irreducible (1 PI) diagrams. \cref{reg1,reg2} represent the two-point vertex function at different orders, each composed of 1 PI diagrams. The composition of all 1PI diagrams, excluding  the zeroth-order one, is termed the self-energy and is often represented by $\Sigma$ \cite{Kleinert2001,Padmanabhan2016}.

The general structure of two-point one-loop order vertex function is
\begin{equation}
	\begin{aligned}\label{reg1}
		\Gamma^{(2)}=\left( \Twopointsimple \right)^{-1}-\Sigma^{(1)}
	\end{aligned}
\end{equation}
where $\Sigma^{(1)}$ is the self-energy at the one loop approximation with
\begin{equation}\label{selfenergya0}
	\Sigma^{(1)} = \left(\frac{1}{2} \TheTadpole  \right)
\end{equation}
The general structure of two-point two-loop order vertex function is 
\begin{equation}
	\begin{aligned}\label{reg2}
		\Gamma^{(2)}=\left( \Twopointsimple \right)^{-1}-\Sigma^{(2)}
	\end{aligned}
\end{equation}
with
\begin{equation}
	\begin{aligned}\label{selfenergya}
		\Sigma^{(2)}=\left(\frac{1}{2} \TheTadpole+\frac{1}{4} \TheDoublebubble + \frac{1}{6} \TheSunrise \right)
	\end{aligned}
\end{equation}

where $\Sigma^{(2)}$ is the self-energy at the two-loop approximation. As one evaluates the diagrams in \cref{selfenergya0,selfenergya} using dimensional regularization \cite{Kleinert2001,Hooft1972,Hooft1973}, divergences appear. These divergences can be removed by introducing appropriate counterterms defined using the \emph{minimal subtraction scheme} (MS-Scheme). \\

The self-energy diagrams for one and two-loop approximation with counterterms can be expressed as
\begin{equation}\label{regselfenergyone}
	\bar{\Sigma}^{(1)} = \frac{1}{2} \TheTadpole + \masscounterterm + \momentumcounterterm \\
\end{equation}
and
\begin{equation}
	\begin{aligned}\label{regselfenergytwo}
		\bar{\Sigma}^{(2)} = 
			\frac{1}{2} \TheTadpole + \masscounterterm + \momentumcounterterm 
			+ \frac{1}{4} \TheDoublebubble + \frac{1}{2} \countertermtwo 
			+\frac{1}{6} \TheSunrise + \frac{1}{2} \countertermthree
	\end{aligned}
\end{equation}
In \cref{onelooptwopoint,twolooptwopoint}, we follow the dimensional regularization procedure \cite{Kleinert2001} and analyze the self-energy diagrams shown in \cref{regselfenergyone,regselfenergytwo} in both ITF and QFT for one-loop and two-loop orders, respectively, using the approximation \( \lambda = g \mu^\epsilon \). It should be noted that in the context of finite terms, \( \lambda \to g \mu^\epsilon \approx g \) as \( \epsilon \to 0 \).

\subsection{One loop order two point function}\label{onelooptwopoint}
The one-loop order two-point vertex function \cite{Kleinert2001} in general form can be written as 
\begin{equation}
	\begin{aligned}\label{gamma21}
		\Gamma^{(2)}=\left( \Twopointsimple \right)^{-1}-\left(\frac{1}{2} \TheTadpole  \right)
	\end{aligned}
\end{equation}
However, one could relate the above tadpole diagram in ITF with a similar QFT diagram using \cref{vacuumtad1,vacuumtad2,vacuumtad3} at \( \lambda = g \mu^\epsilon \) as 
\begin{equation}
	\begin{aligned}\label{tadpole_relation}
		\TheTadpole_\itf = \TheTadpole_\qft -g \mu^\epsilon  S_{1}(m,T)
	\end{aligned}
\end{equation}
with
\begin{equation}
	\begin{aligned}
		S_1(m,T)&=\int \frac{n_B(\beta \varepsilon_p)}{\varepsilon_p} \frac{\dd[3]p}{(2 \pi)^3}
		=\frac{1}{\pi} \sum_{n=1}^\infty \left( \frac{m}{2 \pi n \beta} \right)K_1(n \beta m)
	\end{aligned}
\end{equation}
(See \cref{diagram1} for detailed derivation).\\
Thus, it is evident that the divergence that comes in the ITF diagram shown in \cref{tadpole_relation} at the one-loop order originates from the corresponding diagram in QFT, or in other words, the divergence parts of both of these diagrams are the same. It is clear from \cref{vacuumtad2} corresponding to the QFT diagram that the integral is diverging. On substituting $\lambda=g\mu^\epsilon$ and $\frac{\dd[4]{p}}{(2 \pi)^4} \to \frac{\dd[4-\epsilon]{p}}{(2 \pi)^{4-\epsilon}}$, we can isolate the term which diverges at the limit \(\epsilon \rightarrow 0\) \cite{Kleinert2001}.
Therefore, the analytical expression can be written up to the first order \(\epsilon\) as
\begin{equation}
	\begin{aligned}
		\TheTadpole_\qft  &= \frac{m^2g}{(4 \pi)^2}\left[ \frac{2}{\epsilon}+ \clubsuit_2(m,\mu)  + \mathcal{A}(\epsilon) \right] +\mathcal{O}(\epsilon^2)
	\end{aligned}
\end{equation}
where
\begin{equation}
	\begin{aligned}
		&		 \frac{			\mathcal{A}(\epsilon)}{\epsilon} = \frac{1}{4} \bigg[  {{[\clubsuit_2} (m,\mu)]^2} + \frac{\pi^2}{3} - \psi'(2) \bigg] 
	\end{aligned}
\end{equation}
\begin{equation}\label{clubsuit}
	\clubsuit_n(m,\mu) = \psi(n)+\ln \left( \frac{4 \pi \mu^2}{m^2} \right) 
\end{equation}
with
\begin{equation}
	\psi(n)=-\gamma_\text{E}+\sum_{l=1}^{n-1} \frac{1}{l} \text{   and   } \psi'(n)=\frac{\pi^2}{6}-\sum_{l=1}^{n-1} \frac{1}{l^2}
\end{equation}
$\gamma_\text{E}$ is \emph{Euler's constant}. \\ \\
To make the two-point vertex function finite, counterterms can be defined to cancel the divergences \cite{Kleinert2001}. 
$\masscounterterm$ and $\momentumcounterterm$ are called mass counterterm and momentum counterterm respectively \cite{Kleinert2001}.
The counterterms can be determined using the pole-picking operator, denoted as \( \mathcal{K} \), where \( \mathcal{K}(A \epsilon^{-n} + B \epsilon^l + C) = A \epsilon^{-n} \) for \( l, n > 0 \).
Therefore
\begin{equation}
	\begin{aligned}\label{counterfirstorder}
		\masscounterterm =-m^2c_{m^2}^1	&=-\frac{1}{2} \mathcal{K} \left( \TheTadpole_\itf \right) =-\frac{1}{2} \mathcal{K} \left( \TheTadpole_\qft \right) =-\frac{m^2g}{(4\pi)^2} \frac{1}{\epsilon} 
	\end{aligned}
\end{equation}
i.e., at the one loop order $\masscounterterm_\qft=\masscounterterm_\itf $.
For the sake of completeness, it is important to note that the momentum counter term at one-loop order is zero.
\begin{equation}
	\momentumcounterterm =-K^2c_\phi^1 = 0
\end{equation}
The general finite vertex function at one-loop order is thus
\begin{equation}
	\begin{aligned}\label{gammafirstorder}
		\bar{\Gamma}^{(2)}= \left( \Twopointsimple \right)^{-1}-\left(\frac{1}{2} \TheTadpole + \masscounterterm  \right) = \text{Finite}
	\end{aligned}
\end{equation}
The specific versions for $\itf$ and $\qft$ can be written as 
\begin{equation}
	\begin{aligned}\label{tadone}
		\bar{\Gamma}^{(2)}_\itf = \left( \Twopointsimple_\itf \right)^{-1}-\left(\frac{1}{2} \TheTadpole_\itf + \masscounterterm_\itf  \right) 
	\end{aligned}
\end{equation}
\begin{equation}\label{tadtwo}
	\bar{\Gamma}^{(2)}_\qft = \left( \Twopointsimple_\qft \right)^{-1}-\left(\frac{1}{2} \TheTadpole_\qft + \masscounterterm_\qft  \right) 
\end{equation}
At the zero external momentum limit, or at \(\Twopointsimple_\itf = \Twopointsimple_\qft\), the two-point one-loop order finite vertex function in ITF and QFT can be connected using \cref{counterfirstorder,gammafirstorder,tadpole_relation,tadone,tadtwo} as 
\begin{equation}
	\begin{aligned}
		\bar{\Gamma}^{(2)}_\itf = \bar{\Gamma}^{(2)}_\qft + \frac{g \mu^\epsilon}{2}S_1(m,T)
	\end{aligned}
\end{equation}
.
\subsection{Two-loop order two point functions}\label{twolooptwopoint}
At the two-loop approximation, the first-order counter term defined earlier to make the one-loop finite vertex function can produce some extra diagrams (referred to as counterterm diagrams) in addition to the second-order diagrams. i.e.,
\begin{equation}
	\begin{aligned}\label{gamma22}
		\bar{\Gamma}^{(2)}=\left( \Twopointsimple \right)^{-1} &- \left(\frac{1}{2} \TheTadpole + \masscounterterm + \momentumcounterterm + \frac{1}{4} \TheDoublebubble + \frac{1}{2} \countertermtwo + \frac{1}{6} \TheSunrise + \frac{1}{2} \countertermthree \right)
	\end{aligned}
\end{equation}

We use the following notations:  \begin{equation}
	\begin{aligned}
		&\sumint_{p}=\sum_{n_p=-\infty}^\infty \int \DD{3}{p} , \varepsilon_p=\sqrt{p^2+m^2}, \omega_{n_p}=2\pi n_p T \text{ , and } n_B(x)=\left[\exp(x)-1 \right]^{-1} \text{.}
	\end{aligned}
\end{equation}
The diagrams in \cref{selfenergya,gamma22}, aside from the tadpole diagram already explained in \cref{vacuumtad1,vacuumtad2,vacuumtad3}, can be expressed mathematically in ITF and QFT as follows:
\begin{align}
	\TheDoublebubble_\itf =  \sumint_{p_1,p_2} \greenna{\lambda^2T^2}{p_1} {\greenn{p_2}}^2 
\end{align}
\begin{align}
	\TheDoublebubble_\qft =  \int \frac{\lambda^2}{\varepsilon_{P_1}^2} \left( \frac{1}{\varepsilon_{P_2}^2} \right)^2 \DD{4}{P_1} \DD{4}{P_2} 
\end{align}
At the limit \( \lambda \to g \mu^\epsilon \), $\frac{\dd[4]{P}}{(2 \pi)^4} \to \frac{\dd[4-\epsilon]{P}}{(2 \pi)^{4-\epsilon}}$ and $\frac{\dd[3]{p}}{(2 \pi)^3} \to \frac{\dd[3-\epsilon]{p}}{(2 \pi)^{3-\epsilon}}$, the diagrams in ITF and QFT can be connected as 

\begin{equation}
	\begin{aligned}\label{doublebub}
		\frac{1}{4} \TheDoublebubble_\itf = &\frac{1}{4} \TheDoublebubble_\qft -\frac{g \mu^\epsilon S_0(m,T)}{16 \pi}   \TheTadpole_\qft  
		+ \frac{ g \mu^\epsilon S_1(m,T)}{4} \pdv{m^2}  \left[ \TheTadpole_\qft  \right] \\
		&+\frac{g^2 \mu^{2 \epsilon}}{16 \pi}S_1(m,T)S_0(m,T)
	\end{aligned}
\end{equation}
where
\begin{equation}\label{SNmT}
	S_N(m, T)=\left( \frac{1}{\pi} \right)\sum_{n=1}^\infty \left( \frac{m}{2\pi n \beta} \right)^N K_N(n m \beta)
\end{equation}

A detailed derivation of \cref{doublebub} with necessary steps can be found in \cref{diagram2}.

The expression for $\TheSunrise$ in ITF and QFT are 
\begin{align}
	\sumint_{p_1,p_2} \frac{1}{\varepsilon_{p_1}^2+\omega_{n_{p_1}}^2} \frac{\lambda^2}{\varepsilon_{p_2}^2+\omega_{n_{p_2}}^2} \frac{T^2}{\varepsilon_{p_1+p_2+s}^2+\omega_{n_{p_1+p_2+s}}^2} 
\end{align}
and 
\begin{align}
	\int \frac{\lambda^2}{\varepsilon_{P_1}^2} \frac{1}{\varepsilon_{P_2}^2} \frac{1}{\varepsilon_{P_1+P_2+S}^2} \DD{4}{P_1} \DD{4}{P_2}
\end{align} 
respectively. At the limit of zero external momentum ($S$=0), the diagram $\TheSunrise$ in ITF and  QFT can be linked ($\lambda \to g \mu^\epsilon$) as
\begin{equation} \begin{aligned}\label{sunr}
		\frac{1}{6}\TheSunrise_{\substack{\itf \\ S=0}} =& \frac{1}{6} \TheSunrise_{\substack{\qft\\S=0}} +  \frac{g^2m^2}{64 \pi^4}  Y(m,T)
		+ \frac{g^2 S_1(m,T)}{2(4\pi)^2} \left( \clubsuit_1(m,\mu) +2-\frac{\pi}{\sqrt{3}} \right) \\
		\\ &+ \frac{g  S_1(m,T)}{2} \pdv{m^2} \mathcal{K} \left( \TheTadpole_\qft \right) \\
\end{aligned} \end{equation} 
where
\begin{equation}
	\begin{aligned}\label{Ymt}
		Y(m,T) = \int_0^\infty \int_0^\infty U(x) U(y) G(x,y) \ \dd x \ \dd y
	\end{aligned}
\end{equation}
with
\begin{equation}
	\begin{aligned}
		&U(x)=\frac{\sinh(x)}{\exp \left( \beta m \cosh(x) \right)-1}
	\end{aligned}
\end{equation}
\begin{equation}
	\begin{aligned}
		&G(x,y)=\ln \left( \frac{1+2 \cosh(x-y)}{1+2 \cosh(x+y)} \frac{1-2 \cosh(x+y)}{1-2 \cosh(x-y)} \right) 
	\end{aligned}
\end{equation}
(Refer to \cref{A-4} for a detailed derivation).

The counterterm diagrams can also be connected in a similar manner.
\begin{equation}
	\begin{aligned}\label{count1}
		\frac{1}{2} \countertermtwo_\itf &= \frac{1}{2} \countertermtwo_\qft 
		+\frac{g}{16 \pi} S_0(m,T) \mathcal{K}\left[  \TheTadpole_\qft \right]
	\end{aligned}
\end{equation}
The comprehensive derivation of the counterterm illustrated in \cref{count1}, including all essential steps, is presented in \cref{counterterm2}.
\\
The remaining counterterm $\bigg(\countertermthree\bigg)$ diagram's  ITF and QFT relation from \cref{counter3} is
\begin{equation}
	\begin{aligned}\label{thirdcount}
		\frac{1}{2} \countertermthree_\itf &= \frac{1}{2} \countertermthree_\qft 
		-\frac{3gS_1(m,T)}{4} \pdv{m^2} \mathcal{K} \left(\TheTadpole_\qft \right)
	\end{aligned}
\end{equation}
By combining the relations from \cref{tadpole_relation,doublebub,count1,sunr,thirdcount}  and applying them to \cref{gamma22}, at the limit \(\lambda \to g \mu^\epsilon\) as \(\epsilon \to 0\), it becomes clear that the divergence of the non-finite two-point vertex function at the two-loop order for ITF is identical to that of QFT.
\begin{equation}
	\begin{aligned}
		\mathcal{K} \left( \bar{\Gamma}^{(2)}_\itf \right) = \mathcal{K} \left( \bar{\Gamma}^{(2)}_\qft \right)
	\end{aligned}
\end{equation}
Combining this with \cref{reg2} also implies
\begin{equation}
	\begin{aligned}
		\mathcal{K} \left( \bar{\Sigma}^{(2)}_\itf \right) = \mathcal{K} \left( \bar{\Sigma}^{(2)}_\qft \right)
	\end{aligned}
\end{equation}
The vertex function can be made finite by subtracting the poles term, which can be determined by applying the $\mathcal{K}$ operator, i.e.,
\begin{equation}
	\begin{aligned}
		\bar{\Gamma}_\itf^\text{(2)finite} &= \bar{\Gamma}^{(2)}_\itf - \mathcal{K} \left( \bar{\Gamma}^{(2)}_\itf \right) \\
		\bar{\Gamma}_\qft^\text{(2)finite} &= \bar{\Gamma}^{(2)}_\qft - \mathcal{K} \left( \bar{\Gamma}^{(2)}_\qft \right)
	\end{aligned}
\end{equation}
In order to simplify further calculations, let us define a new operator \(\Delta\) such that \begin{equation}
	\begin{aligned}\label{newoperator}
		\Delta (\mathbb{A}) &=\mathbb{A}_\itf|_{k,\omega_{n_k}=0}-\mathcal{K} \left( \mathbb{A}_\itf\right)|_{k,\omega_{n_k}=0} 
		-\mathbb{A}_\qft|_{k_0,k=0}+\mathcal{K} \left( \mathbb{A}_\qft \right)|_{k_0,k=0} 
\end{aligned} \end{equation} where $\mathbb{A}$ represents the appropriate diagram. 

Using this new notation, the finite vertex functions can be correlated under SMC as shown below.
\begin{equation}
	\begin{aligned}\label{thediff}
		\bar{\Gamma}_\itf^\text{(2)finite}=\bar{\Gamma}_\qft^\text{(2)finite} + \bar{\Gamma}^\text{(2)diff}
	\end{aligned}
\end{equation}
where
\begin{equation}
	\begin{aligned}\label{a1}
		-\bar{\Gamma}^\text{(2)diff} &= \frac{\Delta}{2} \left( \TheTadpole \right) + \frac{\Delta}{4} \left( \TheDoublebubble \right) + \frac{\Delta}{6} \left( \TheSunrise \right) 
		=\Delta \Sigma^{\text{(2)}}
	\end{aligned}
\end{equation}
with
\begin{equation}
	\begin{aligned}\label{a2}
		\frac{1}{2} \Delta \left( \TheTadpole \right)& =-\frac{g}{2}S_1(m,T) 
	\end{aligned}
\end{equation}
\begin{equation}
	\begin{aligned}\label{a3}
		\frac{1}{4} \Delta \left( \TheDoublebubble \right)=&\frac{g^2}{16 \pi}S_0(m,T)S_1(m,T) 
		-\frac{g^2 m^2}{4 (4\pi)^3}S_0(m,T) \left[\clubsuit_2(m,\mu) \right] \\
		&+\frac{g^2}{4(4 \pi)^2}S_1(m,T) \left[\clubsuit_1(m,\mu) \right] 
	\end{aligned}
\end{equation}
\begin{equation}
	\begin{aligned}\label{a4}
		\frac{1}{6}\Delta \left( \TheSunrise \right) =& \frac{g^2}{2(4 \pi)^2}S_1(m,T) \left[ \clubsuit_1(m,\mu)   \right] 
		+ \frac{g^2}{2(4 \pi)^2}S_1(m,T)  \left( 2-\frac{\sqrt{3} \pi}{3} \right)\\
		&+\frac{g^2m^2}{64 \pi^4}  Y(m,T)
	\end{aligned}
\end{equation}
In reference \cite{Arjun2022}, it was demonstrated that the divergence factors of two-point and four-point vertex functions for both ITF and QFT were in the same form. Likewise, the renormalization group functions, such as beta ($\beta(g)$) and gamma ($\gamma_m(g)$), exhibited similar formats for both ITF and QFT approximations. Subsequently, in \cite{Arjun2022}, the renormalization group equation was simultaneously applied to both vertex functions in ITF and QFT. The resulting expression established a new coupling constant relation. In other words, when the RGE equations were applied and approximated as 
\begin{equation}
	\rge \ \bar{\Gamma}^{(2)}_\itf \approx 0
\end{equation} and 
\begin{equation}
\rge \ \bar{\Gamma}^{(2)}_\qft \approx 0
\end{equation}, they resulted in 
\begin{equation}
	 \rge \ \bar{\Gamma}^\text{(2)diff} \approx 0
\end{equation}. Upon solving this with the assumption $\bar{\Gamma}^\text{(2)diff} \neq 0$, we obtained a new coupling constant relation in addition to the beta and gamma function relations. These relations were simultaneously solved in reference \cite{Arjun2022} under the same mass scale and coupling approximation, and the running mass and running coupling constant were computed numerically.
\section{Self-energy Approximation}\label{selfenergy}
The other possibility, apart from \cite{Arjun2022}, where $\rge \ \bar{\Gamma}^\text{diff}$ becomes zero is when $\bar{\Gamma}^\text{diff}$ itself is zero. By setting $\bar{\Gamma}^\text{diff}$ to zero and combining \cref{thediff,reg2}, we find that 
\begin{equation}
	\bar{\Sigma}^{(2)}_\itf = \bar{\Sigma}^{(2)}_\qft \text{ .}
\end{equation}
That is, when the self-energy function of both ITF and QFT becomes equal. Therefore 
\begin{equation}
	\begin{aligned}\label{gammadiffzero}
		&-\bar{\Gamma}^\text{(2)diff} = \Delta \Sigma^{(2)} = 0 
	\end{aligned}
\end{equation}
Combining \cref{gammadiffzero} with \cref{a1,a2,a3,a4} we get the non-trivial expression
\begin{equation} \label{coupling}
	\begin{aligned}
		&g =  \frac{-\frac{1}{2} g^{-1} \Delta \left[ \TheTadpole \right] }{ g^{-2} \left( \frac{1}{4} \Delta \left[ \TheDoublebubble \right] + \frac{1}{6} \Delta \left[ \TheSunrise_{S=0} \right] \right)}=\frac{S_1(m,T)}{  D_{01}(m,T) -D_{0}(m,T,\mu) + D_1(m,T,\mu) } 
	\end{aligned}
\end{equation}
where
\begin{equation}
	\begin{aligned}
		D_{01}(m,T) &=\frac{S_0(m,T)S_1(m,T)}{8 \pi}   + \frac{m^2}{32 \pi^4}  Y(m,T)   \\
		D_{0}(m,T,\mu) &=\frac{m^2}{2 (4\pi)^3}S_0(m,T) \clubsuit_2(m,\mu)\\
		D_{1}(m,T,\mu) &= \left[ \frac{3}{2} \clubsuit_1(m,\mu) + 2 - \frac{\sqrt{3} \pi}{3} \right] \frac{S_1(m,T)}{16 \pi^2} 
	\end{aligned}
\end{equation}
and the expressions for \(\clubsuit_n(m,\mu)\),  \( S_N(m,T) \) and \( Y(m,T) \) are given in \cref{clubsuit,Ymt,SNmT}, respectively. \\

If we combine with beta coupling constant relation \cite{Kleinert2001} such as
\begin{equation}\label{mass-scale0}
	\pdv{g(\mu)}{\ln (\mu)} =  \beta_2 g^2+ \beta_3 g^3
\end{equation}
give rise to the result
\begin{align}\label{mass-scale}
	\ln(\mu/\mu_0) &=\int^{g} \frac{1}{\beta_2 t^2+ \beta_3 t^3} dt  = - \frac{1}{\beta_2 \ g}+ \frac{\beta_3}{\beta_2^2} \ln \left(\beta_3+ \frac{\beta_2}{g} \right)
\end{align}
Similarly, the corresponding running mass coupling relation is
\begin{equation}\label{runningmass}
	\pdv{\ln m}{\ln \mu}=\gamma_m (g)
\end{equation}
When combined with \cref{mass-scale0}, it results in
\begin{equation}
	\begin{aligned}
		&\frac{\partial  \ \ln(m)}{\partial g } \pdv{g}{\ln \mu} =\gamma_m(g) \implies \frac{\partial \ \ln(m)}{\partial g}=\frac{\gamma_m(g)}{\beta(g)} = \frac{\gamma_{m1}+\gamma_{m2} g}{\beta_2 g+ \beta_3 g^2}   \\
		\text{i.e., } &\ln \left( \frac{m}{m_0} \right) =  \frac{\gamma_{m1}}{\beta_2} \ln (g) + \left( \frac{\gamma_{m2}}{\beta_3}-\frac{\gamma_{m1}}{\beta_2} \right)  \ln(\beta_3 \ g+\beta_2) 
		\label{mass-coupling}
	\end{aligned}
\end{equation}
where $\mu_0$, and $m_0$ are the respective integration constants. To simplify the equation further, we choose the integration constants \(\mu_0 = m_0 = 1\). In the zero momentum limit, we choose \(\gamma_{m_1}=\frac{1}{2} \frac{1}{(4 \pi)^2}\), \(\gamma_{m_2}=-\frac{1}{2} \frac{1}{(4 \pi)^4}\),  \(\beta_2= \frac{3}{(4 \pi)^2}\) and \(\beta_3=-\frac{6}{(4 \pi)^4}\) \cite{Arjun2022,Kleinert2001}. \\

Solving \cref{mass-scale,mass-coupling,coupling} simultaneously, we obtain the temperature-dependent running mass, mass scale and coupling constant.

\subsection{Free energy contribution from temperature dependent terms in the vacuum diagrams}
Following the conventions in reference \cite{Kleinert2001}, the zero-point vertex function can be written as
\begin{equation}
	\begin{aligned}
		\Gamma^0=\frac{1}{2} \TheVacuuma + \frac{1}{8} \TheVacuumb
	\end{aligned}
\end{equation}
and the free energy density \cite{Laine2016} as
\begin{equation}
	\begin{aligned}\label{freeenergy1}
		\mathcal{F}=\frac{1}{2} \TheVacuuma_\itf - \frac{1}{8} \TheVacuumb_\itf
	\end{aligned}
\end{equation}
where
\begin{equation}
	\begin{aligned}
		\TheVacuuma_\itf &= T \sum_n \int \frac{\dd[3]p}{(2\pi)^3} \ln \left(\omega_n^2+\va{p}^2+m^2 \right) 
	\end{aligned}
\end{equation}
\begin{equation}
	\begin{aligned}
		\TheVacuumb_\itf &= -\lambda \left( T \sum_n \int \frac{\dd[3]p}{(2\pi)^3} \frac{1}{\omega_n^2+\va{p}^2+m^2} \right)^2
	\end{aligned}
\end{equation}

The expression $\mathcal{F}$ in \cref{freeenergy1} represents the non-finite version of free energy density.
The one-loop vacuum diagram $\TheVacuuma$ stands alone as the only diagram devoid of any vertex.
Following some algebraic manipulations, the thermal and non-thermal diagrams can be connected at the one-loop order as 
\begin{equation}
	\begin{aligned}\label{vaccuma}
		\TheVacuuma_\itf= \TheVacuuma_\qft- 4 \pi S_2(m,T)
	\end{aligned}
\end{equation}
with
\begin{equation}
	\begin{aligned}
		\TheVacuuma_\qft &= \int \frac{\dd[4]P}{(2\pi)^4} \ln \left( P^2+m^2 \right) \\
	\end{aligned}
\end{equation}
See \cref{vacuumdiagram1} for detailed derivation.
As we expand it in the powers of $\epsilon$ \cite{Kleinert2001} we have 
\begin{equation}
	\begin{aligned}
		\TheVacuuma_\qft &= -\frac{m^4}{2 \mu^\epsilon (4\pi)^2}  \left[ \frac{2}{\epsilon} + \clubsuit_2(m,\mu)+\frac{1}{2} \right] + \mathcal{O}(\epsilon) 
	\end{aligned}
\end{equation}
Similarly
\begin{equation}
	\begin{aligned}\label{vacuumb}
		\TheVacuumb_\itf =&	\TheVacuumb_\qft - \lambda S_1^2(m,T) + 2 \left[\lbrace -\lambda \rbrace \TheTadpole_\qft \right] S_1(m,T) 
	\end{aligned}
\end{equation}
The braces $\lbrace \lambda \rbrace$ serve as labels rather than multiplication factors.
As $\lambda \to g \mu^\epsilon$
\begin{equation}
	\begin{aligned}\label{vacuumbtwo}
		\TheVacuumb_\itf=&	\TheVacuumb_\qft - g \mu^\epsilon S_1^2(m,T) +\frac{ 2 m^2g}{(4 \pi)^2}\left[ \frac{2}{\epsilon}+ \clubsuit_2(m,\mu) \right] S_1(m,T) + \mathcal{O}(\epsilon)
	\end{aligned}
\end{equation}
Refer to \cref{vacuumdiagram2} for a comprehensive derivation.
From \cite{Kleinert2001}, we have
\begin{equation}
	\begin{aligned}
		\TheVacuumb_\qft = -\lambda \left(\int \frac{\dd[4]P}{(2\pi)^4} \frac{1}{P^2+m^2} \right)^2 
		=&-\frac{m^4g}{(4\pi)^4 \mu^\epsilon} \left[ \left( \frac{2}{\epsilon}+\clubsuit_2(m,\mu) \right)^2 + 4\frac{\mathcal{A}(\epsilon)}{\epsilon} \right]\\
		&+\mathcal{O}(\epsilon)
	\end{aligned}
\end{equation}

However, if we include the vacuum counter term following the same procedures as in \cite{Kapusta1979,Kapusta2006} and in \cref{vacuumcounterterm}, then the counter terms in thermal and non-thermal QFT can be expressed as
\begin{equation}
	\begin{aligned}\label{vacuumcounterone}
		\TheVacuumCounterterm_\itf=\TheVacuumCounterterm_\qft-\frac{1}{2} \mathcal{K} \left[ \TheTadpole_\qft \right]S_1(m,T)
	\end{aligned}
\end{equation}
where
\begin{align}
	\TheVacuumCounterterm_\itf = -\frac{1}{2} \mathcal{K} \left[ \TheTadpole_\itf \right] \times \frac{\partial}{\partial m^2} \left[ \TheVacuuma_\itf \right]
\end{align}
\begin{align*}
	\TheVacuumCounterterm_\qft = -\frac{1}{2} \mathcal{K} \left[ \TheTadpole_\qft \right] \times \frac{\partial}{\partial m^2} \left[ \TheVacuuma_\qft \right]
\end{align*}
These results also indicate that the two-loop order zero-point vertex function in ITF and non-thermal QFT are having the same order of divergence if counterterms are considered. i.e., the divergence/pole terms of one-loop vacuum diagram for non-thermal QFT and ITF are the same under SMC ( See \cref{vaccuma} ).
\begin{equation}
	\begin{aligned}
		&\frac{1}{2}	\mathcal{K} \left[ \TheVacuuma_\itf \right] = \frac{1}{2} \mathcal{K} \left[ \TheVacuuma_\qft \right] 
	\end{aligned}
\end{equation}
The same trend can be seen in the summation of two-loop vacuum diagrams (\cref{vacuumb}) with the counterterm (\cref{vacuumcounterone}) as shown below
\begin{equation}
	\begin{aligned}
		\text{I}(\epsilon) & =\frac{1}{8} \mathcal{K} \left[ \TheVacuumb_\itf \right] + \frac{1}{2} \mathcal{K} \left[ \TheVacuumCounterterm_\itf \right]  
		=  	\frac{1}{8} \mathcal{K} \left[ \TheVacuumb_\qft \right] + \frac{1}{2} \mathcal{K} \left[ \TheVacuumCounterterm_\qft \right]
	\end{aligned}
\end{equation}
Therefore, at $\lambda = g \mu^\epsilon$, using \cref{vaccuma,vacuumbtwo,newoperator}, we get
\begin{equation}
	\begin{aligned}\label{freeenergy2}
		\frac{1}{2} \Delta \left[ \TheVacuuma_\itf \right] =& -2 \pi S_2(m, T) \\
		\frac{1}{8} \Delta \left[ \TheVacuumb_\itf \right] =& \frac{1}{4} \frac{m^2g}{(4 \pi)^2}S_1(m,T) \clubsuit_2(m,\mu) 
		- \frac{g}{8} S_1^2(m,T)
	\end{aligned}
\end{equation}
Now using \cref{freeenergy1,freeenergy2}
\begin{equation}
	\begin{aligned}\label{freeenergy3}
		\Delta \mathcal{F} =& -2 \pi S_2(m, T) + \frac{g}{8}S_1^2(m,T) 
		- \frac{1}{4} \frac{m^2g}{(4 \pi)^2}S_1(m,T) \clubsuit_2(m,\mu)
	\end{aligned}
\end{equation}
The quantity $\Delta \mathcal{F}$ is finite.
This result corresponds to the two-loop contribution to finite free energy density, while the first term corresponds to the one-loop. This expression matches the free energy expression in \cite{Andersen2000}, albeit with a difference in the choice of coupling (using $g^2$ instead of $g$), along with some minor conventional distinctions. As we substitute the temperature-dependent running mass, running coupling, and mass scale obtained from the previous section into the expression for the finite free energy density, we can determine how the finite free energy density varies with respect to temperature.
Now at limit $ \beta m \to 0$ we can write
\begin{equation}
	\begin{aligned}
		\lim_{\beta m \to 0} S_1(m, T) &= \frac{T^2}{12} \\
		\lim_{\beta m \to 0} S_2(m, T) &= \frac{1}{2\pi} \frac{\pi^2}{90} T^4
	\end{aligned}
\end{equation}
\begin{equation}
	\begin{aligned}
		\lim_{\beta m \to 0}	\Delta \mathcal{F} &= - \frac{\pi^2}{90} T^4 +\frac{g}{4!} \frac{T^4}{48}   - \frac{T^2}{48} \frac{m^2g}{(4 \pi)^2} \left[ \psi(2) + \ln \left( \frac{4 \pi \mu^2}{m^2} \right) \right]
	\end{aligned}
\end{equation}
At the massless limit, $-\Delta \mathcal{F}$ is identical with the corresponding pressure expression for $\phi^4$ theory given in reference \cite{Kapusta2006}. The only distinction lies in our usage of $(\lambda/4!) \phi^4$ theory, whereas in reference \cite{Kapusta2006}, $\lambda \phi^4$ theory is employed, resulting in a factor difference of $4!$ in the pressure expression's terms involving the coupling. Incorporating this multiplicative correction will reconcile both expressions. \\

The free energy density vs. temperature relation becomes particularly interesting when we apply the temperature-dependent running coupling constant, running mass, and mass scale derived from the previous section. When applied here, it provides the temperature vs. free energy density relation, as shown in \cref{fig2}, which is novel. This is because the coupling constant is computed without relying on multiple RGEs, but rather by simply applying the already established RGE equations to both thermal and non-thermal vertex functions simultaneously.
\subsection{Pressure evaluation using quasi-particle model}
Several phenomenological models have been proposed to explain the non-ideal behavior of Quark-Gluon Plasma (QGP) observed in lattice QCD simulations and heavy-ion collision experiments. These include the strongly interacting QGP (sQGP) model \cite{Shuryak2005}, in which the presence of various neutral and colored bound states, arising from Coulomb interactions, leads to non-ideal effects. Another significant approach is the strongly coupled QGP (SCQGP) model, which draws an analogy between QGP near the critical temperature \( T_c \) and a strongly coupled plasma (SCP) in QED, employing a modified equation of state to better align QCD lattice data \cite{Bannur2006}. Among these, the quasiparticle QGP (qQGP) model has emerged as one of the most successful frameworks, describing QGP as a medium composed of quasiparticles with temperature-dependent effective masses. Although initial formulations \cite{Goloviznin1993,Peshier1994} encountered thermodynamic inconsistencies, subsequent refinements—such as rederiving the model from the energy density—have resolved these issues \cite{Bannur2007a}, making quasiparticle models a robust tool for investigating the thermodynamic properties of QGP. In this section, we extend this approach to the $\phi^4$ theory, comparing the resultant pressure with the free energy density. \\

According to the quasiparticle model in \cite{Bannur2007,Bannur2007a,Bannur2007b}, the quasiparticle mass per temperature is proportional to the coupling constant i.e., $m_q/T \propto g(T)$.
The energy density and pressure of a particle obeying the Bose-Einstein distribution can be derived as
\begin{equation}
	\begin{aligned}\label{energydensity}
		\varepsilon(T) = g_f \int \DD{3}{p} \frac{\varepsilon_p}{\exp \left(\beta \varepsilon_p \right)-1} 
		&=g_f \frac{ m_q^4}{2 \pi^2} \sum_{n=1}^\infty \left[  \frac{3K_2(\frac{nm_q}{T})}{(\frac{nm_q}{T})^2}+\frac{K_1(\frac{nm_q}{T})}{\frac{nm_q}{T}} \right] \\
		&=g_f \left[ 6 \pi S_2(m_q, T) + m_q^2 S_1(m_q,T)  \right]
	\end{aligned}
\end{equation}
where $K(n, x)$ is the modified Bessel function of the second kind, and   \\
\(	S_N(m, T)=\left( \frac{1}{\pi} \right)\sum_{n=1}^\infty \left( \frac{m}{2\pi n \beta} \right)^N K_N(n m \beta) \).
\\
We have energy density - pressure relation as
\begin{align}\label{pressure}
	\frac{P}{T}-\frac{P_0}{T_0}=\int_{T_0}^T \frac{\varepsilon(T)}{T^2} \ dT
\end{align}
We define $m_q(T) = g(T) T/(6 \pi)$ following the quasi-particle framework \cite{Bannur2007,Bannur2007a,Bannur2007b}, and select the parameters $T_0=1.0$ and $P_0=0.0$,$0.039$. Subsequently, we obtain a quasiparticle pressure as depicted in \cref{fig2}. Notably, different choices of $P_0$ and $T_0$ values can be considered. These choices may be informed by lattice data, once such data becomes available. 
	\section{Results and Discussion}\label{results}
\begin{figure}[h]
	\includegraphics[scale=1.3]{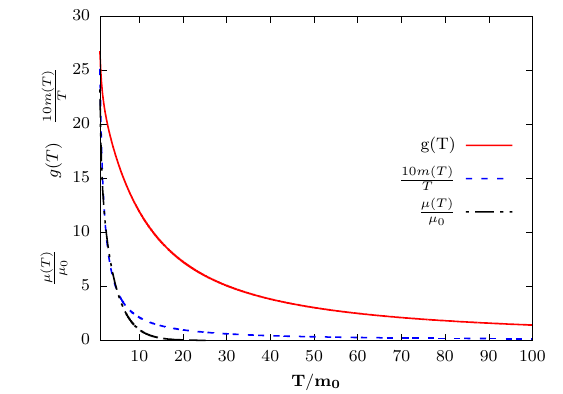}
	\captionof{figure}{The coupling constant, along with the \(\mu(T)\) and scaled running mass per temperature, is plotted against different temperatures. In this figure, \(\mu_0\) and \(m_0\) are approximated to unity.\label{fig1}}
\end{figure}
\begin{figure}[h]
	\includegraphics[scale=1.3]{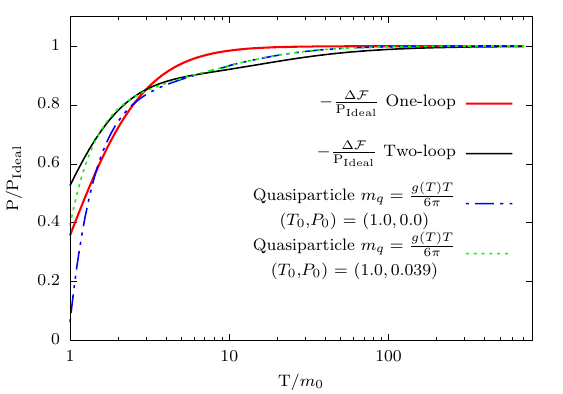}
	\captionof{figure}{The Pressure divided by the ideal pressure $\pi^2T^4/90$, is plotted against different temperatures. In this figure, \(\mu_0\) and \(m_0\) are approximated to unity.\label{fig2}}
\end{figure}
\begin{figure}[h]
	\includegraphics[scale=1.3]{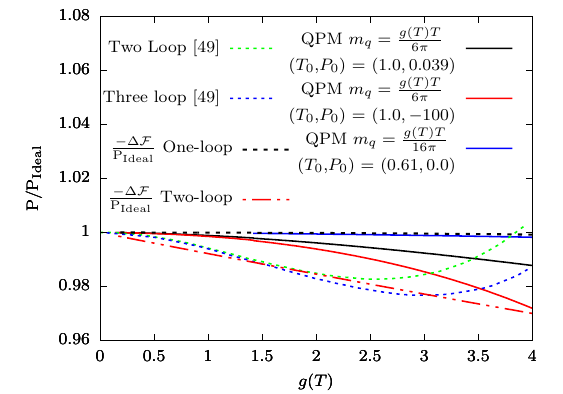}
	\captionof{figure}{The Pressure divided by the ideal pressure $\pi^2T^4/90$, is plotted against $g(T)$. The results are compared with \cite{Andersen2009}.\label{fig3}}
\end{figure}

In this study, we revisit reference \cite{Arjun2022} and compute the running mass, mass scale, and running coupling constant for the massive $\phi^4$ theory within the framework of the self-energy approximation. Within this approximation, we impose the condition that the self-energy calculated at the zero external momentum limit in the Imaginary Time Formalism (ITF) equals that of non-thermal Quantum Field Theory (QFT) under the same mass scale and coupling approximation (SMC). This finding is consistent with existing renormalization group equations. It has been previously demonstrated in \cite{Arjun2022}, that the Renormalization Group (RG) functions take the same form for both QFT and ITF under SMC. \\
By simultaneously solving the renormalization group functions $\beta(g)$ and $\gamma_m(g)$, as represented by \cref{mass-coupling,mass-scale}, with the self-energy approximation expression for $g$ in \cref{coupling}, we obtain temperature-dependent solutions for the running mass,  running coupling constant, and mass scale. \\ \\

We present the main result in  \cref{fig1,fig2}, focusing on the simple approximation of zero external momentum limit.  In \cref{fig1}, we plot scaled functions of the two-loop coupling constant, running mass per temperature, and mass scale on the y-axis as a multi-variable plot. This approach is chosen to provide a comprehensive visualization of the relationships between various quantities as functions of temperature. From \cref{fig1}, it becomes apparent that there exists a negative correlation between the running mass and temperature, as well as between the running coupling and temperature. Additionally, a negative correlation is observed between the mass scale and temperature. These findings indicate that as temperature increases, the values of the running mass, running coupling, and mass scale tend to decrease, although with varying degrees of change. \\

Thermodynamic quantities such as pressure and free energies are closely related \cite{Kapusta2006}. It is well-known that at high temperature limits, the system pressure should approach the ideal Stefan-Boltzmann limit. This provides a useful tool for validating the running mass, coupling constant, and mass scale. Various approaches and models exist for determining thermodynamic quantities. One famous model is the quasiparticle model, where the energy density is calculated by integrating the relativistic energy function over the Bose-Einstein distribution \cite{Bannur2007,Bannur2007a,Bannur2007b}. Integrating the energy density per temperature squared with respect to temperature yields the pressure per temperature, as shown in \cref{pressure}. This pressure relation depends on the running coupling constant and temperature. Applying the running coupling constant-mass relation ($m_q/T=\frac{g(T)}{6\pi}$) to the equation \cref{energydensity} yields the plot shown in \cref{fig2}, where it is evident that the pressure reaches the ideal limit $(\pi^2/90)T^4$ as temperature increases. \\

\cref{fig3} compares the pressure calculated using the quasiparticle model (QPM) and the free energy density calculated using the SEA with the massless \( \phi^4 \) theory under the effective field theory framework \cite{Andersen2009}. At low values of the running coupling constant \( g \), both models predict similar ratios of pressure to ideal pressure. However, as \( g \) increases, their distinct approaches lead to increasingly different results. This difference is understandable, as the analysis in \cite{Andersen2009} employs weak coupling to derive the pressure. The QPM depends on the initial pressure parameters \( T_0 \) and \( P_0 \). The figure includes plots for various combinations of \( T_0 \) and \( P_0 \) to illustrate the dependence of QPM results on these parameters. Since the effective mass in the QPM is proportional to \( g(T)T \), we have also plotted results using different proportionality constants for comparison. To determine the most accurate proportionality constant and the optimal values of \( P_0/T_0 \), lattice data will be crucial. The free energy density calculated using the SEA approach is a function of the running mass, running coupling constant, and mass scale. However, the relationship between \( g \) and free energy density may not fully capture the essence of the system. In this context, a comparison of temperature (\( T \)) versus free energy density may be more relevant. This discussion naturally leads us to Figure 2, where we compare the free energy and pressure against temperature using a similar approach for the coupling constant. \\

In \cref{fig2} the approach involves computing free energy relations using vacuum diagrams. In contrast to the quasiparticle model, which primarily relies on the running coupling constant, the finite free energy density at the one-loop approximation heavily depends on the running mass. However, the free energy density derived from two-loop vacuum diagrams is influenced by the running mass, coupling constant, and mass scale. Therefore, it is a preferable method for assessing the validity of all three parameters. In this work, we have expressed the finite free energy density using the same conventions that were employed to derive the coupling constant. It should be noted that the quasiparticle model, utilizing the mass approximation $m_q(T)=g(T) T / (6 \pi)$, at $(P_0,T_0)=(0.039,1.0)$, closely aligns with the negative of the two-loop free energy density near low-temperature and very high-temperature limits. However, the absence of lattice data hinders a robust parameter fitting process. Nevertheless, the comparison underscores the quasiparticle model's flexibility, particularly through parameters such as the proportionality constant connecting mass per temperature with the coupling constant and the integration constant parameters $P_0$, distinguishing it from other models. As observed in the \cref{fig2}, for both one and two-loop approximations, the negative of the finite free energy density reaches the ideal Stefan-Boltzmann limit at high temperatures. \\
\vspace{ 1 cm}
\section{Conclusion}

In this work, we have demonstrated that it is possible to derive the running mass and running coupling constant consistently without relying on multiple sets of renormalization group equations. A simultaneous application of the existing RGE equation to the imaginary time formalism and non-thermal quantum field theory is sufficient to derive the running mass and coupling constant. Our approach offers a more straightforward method for studying the running coupling constant, effective mass, and mass scale.\\

Here we employ a self-energy approximation (SEA) at zero external momentum limit, resulting in a trivial solution for the running coupling constant $(g)$, running mass $(m)$, and mass scale $(\mu)$. In SEA under Same Mass Scale and Coupling approximation (SMC), the self-energy of non-thermal Quantum Field Theory (QFT) and Imaginary Time Formalism (ITF) are considered equal, aligning with the renormalization group equation. This also implies that in SEA, the finite vertex function for both ITF and non-thermal QFT becomes the same, leading to an additional relation connecting temperature with the coupling constant, mass scale, and running mass, alongside the renormalization group functions $\beta(g)$ and $\gamma_m(g)$ (\cref{coupling,mass-scale0,mass-scale,runningmass,mass-coupling}). Solving these relations simultaneously yields temperature-dependent running mass, mass scale, and coupling function. \\

The pressure and free energy density are calculated using different formulations. To calculate the pressure, we employed the quasiparticle model, where the pressure is a function of the coupling constant and temperature. The free energy density is determined by a function derived from vacuum diagrams involving the running mass, coupling constant, and mass scale. Evaluation has shown that these quantities reach their ideal limits at higher temperatures. Once lattice data or experimental data becomes available within a certain range, it would be intriguing to adjust various integration constants and compare the outcomes. Additionally, extending this model to Quantum Electrodynamics (QED) and Quantum Chromodynamics (QCD) could provide valuable insights into their behavior \cite{Arjun2024}.

\bibliographystyle{JHEP}
\bibliography{main}

\providecommand{\href}[2]{#2}\begingroup\raggedright\begin{thebibliography}{10}

\bibitem{Arjun2022}
K.~Arjun, A.~M. Vinodkumar, and V.~M. Bannur, {\it Running coupling constant in
  thermal ${\ensuremath{\phi}}^{4}$ theory up to two loop order},  {\em Phys.
  Rev. D} {\bf 105} (Jan, 2022) 025023.

\bibitem{Kapusta2006}
J.~I. Kapusta and C.~Gale, {\em Finite-Temperature Field Theory}.
\newblock Cambridge University Press, 2006.

\bibitem{Khanna2009}
F.~C. Khanna, ed., {\em Thermal quantum field theory}.
\newblock World Scientific Pub. Co, Singapore, 2009.
\newblock Includes bibliographical references (p. 445-455) and index.

\bibitem{Millington2014}
P.~Millington, {\em Thermal Quantum Field Theory and Perturbative
  Non-Equilibrium Dynamics}.
\newblock Springer International Publishing, 2014.

\bibitem{Sarkar2010}
S.~Sarkar, H.~Satz, and B.~C. Sinha, eds., {\em The Physics of the Quark-Gluon
  Plasma: Introductory Lectures}.
\newblock No.~785. Springer Berlin Heidelberg, 2010.

\bibitem{Padmanabhan2016}
T.~Padmanabhan, {\em Quantum Field Theory}.
\newblock Springer International Publishing, 2016.

\bibitem{Yagi2005a}
K.~Yagi, T.~Hatsuda, and Y.~Miake, {\em Quark-gluon plasma: From Big Bang to
  Little Bang}.
\newblock Cambridge Univ. Press, 2005.

\bibitem{Bannur2007}
V.~M. Bannur, {\it Revisiting the quasi-particle model of the
  quark{\textendash}gluon plasma},  {\em Eur. Phys. J C} {\bf 50} (feb, 2007)
  629--634.

\bibitem{Bannur2007a}
V.~M. Bannur, {\it Comments on quasiparticle models of quark{\textendash}gluon
  plasma},  {\em Phys. Lett. B} {\bf 647} (apr, 2007) 271--274.

\bibitem{Bannur2007b}
V.~M. Bannur, {\it Self-consistent quasiparticle model for quark-gluon plasma},
   {\em Phys. Rev. C} {\bf 75} (apr, 2007) 044905.

\bibitem{Dyson1949}
F.~J. Dyson, {\it The $s$ matrix in quantum electrodynamics},  {\em Phys. Rev.}
  {\bf 75} (Jun, 1949) 1736--1755.

\bibitem{Stueckelberg1953}
{Stueckelberg, E.C.G.} and {Petermann, A.}, {\it La normalisation des
  constantes dans la théorie des quanta},  {\em Helv. phys. acta} (1953).

\bibitem{GellMann1954}
M.~Gell-Mann and F.~E. Low, {\it Quantum electrodynamics at small distances},
  {\em Phys. Rev.} {\bf 95} (Sept., 1954) 1300--1312.

\bibitem{Bogoliubow1957}
N.~N. Bogoliubow and O.~S. Parasiuk, {\it Über die multiplikation der
  kausalfunktionen in der quantentheorie der felder},  {\em Acta Math.} {\bf
  97} (1957), no.~0 227--266.

\bibitem{Hepp1966}
K.~Hepp, {\it Proof of the bogoliubov-parasiuk theorem on renormalization},
  {\em Commun. Math. Phys.} {\bf 2} (Dec., 1966) 301--326.

\bibitem{Hooft1973}
G.~'t~Hooft, {\it Dimensional regularization and the renormalization group},
  {\em Nucl. Phys. B} {\bf 61} (sep, 1973) 455--468.

\bibitem{Hooft1972}
G.~'t~Hooft and M.~Veltman, {\it Regularization and renormalization of gauge
  fields},  {\em Nucl. Phys. B} {\bf 44} (jul, 1972) 189--213.

\bibitem{Weinberg1973}
S.~Weinberg, {\it New approach to the renormalization group},  {\em Phys. Rev.
  D} {\bf 8} (Nov., 1973) 3497--3509.

\bibitem{Bollini1972}
C.~G. Bollini and J.~J. Giambiagi, {\it Dimensional renorinalization : The
  number of dimensions as a regularizing parameter},  {\em Il Nuovo Cimento B
  (1971-1996)} {\bf 12} (Nov, 1972) 20--26.

\bibitem{Callan1970}
C.~G. Callan, {\it Broken scale invariance in scalar field theory},  {\em Phys.
  Rev. D} {\bf 2} (oct, 1970) 1541--1547.

\bibitem{Symanzik1970}
K.~Symanzik, {\it Small distance behaviour in field theory and power counting},
   {\em Commun. Math. Phys.} {\bf 18} (sep, 1970) 227--246.

\bibitem{Kleinert2001}
H.~Kleinert and V.~Schulte-Frohlinde, {\em Critical Properties of
  Phi4-Theories}.
\newblock {World} {Scientific}, jul, 2001.

\bibitem{Matsubara1955}
T.~Matsubara, {\it A new approach to quantum-statistical mechanics},  {\em
  Prog. Theor. Phys.} {\bf 14} (Oct., 1955) 351--378.

\bibitem{Kapusta1979}
J.~I. Kapusta, {\it Infrared properties of quark gas},  {\em Phys. Rev. D} {\bf
  20} (aug, 1979) 989--995.

\bibitem{Keldysh1964}
L.~Keldysh {\em Sov. Phys. JETP} {\bf 47} (1965) 1515--1527.

\bibitem{Schwinger1961}
J.~Schwinger, {\it Brownian motion of a quantum oscillator},  {\em J. Math.
  Phys.} {\bf 2} (May, 1961) 407--432.

\bibitem{Mahanthappa1962}
K.~T. Mahanthappa, {\it Multiple production of photons in quantum
  electrodynamics},  {\em Phys. Rev.} {\bf 126} (Apr., 1962) 329--340.

\bibitem{Bakshi1963}
P.~M. Bakshi and K.~T. Mahanthappa, {\it Expectation value formalism in quantum
  field theory. ii},  {\em J. Math. Phys.} {\bf 4} (Jan., 1963) 12--16.

\bibitem{CaronHuot2009}
S.~Caron-Huot, {\it Hard thermal loops in the real-time formalism},  {\em J.
  High Energy Phys.} {\bf 2009} (Apr., 2009) 004--004.

\bibitem{Santos2020}
A.~F. dos Santos and F.~C. Khanna, {\it Bhabha scattering in very special
  relativity at finite temperature},  {\em Eur. Phys. J. C} {\bf 80} (Aug.,
  2020).

\bibitem{Lundberg2021}
T.~Lundberg and R.~Pasechnik, {\it Thermal field theory in real-time formalism:
  concepts and applications for particle decays},  {\em Eur. Phys. J. A} {\bf
  57} (Feb., 2021).

\bibitem{Carrington1999}
M.~Carrington, H.~Defu, and M.~Thoma, {\it Equilibrium and non-equilibrium hard
  thermal loop resummation in the real time formalism},  {\em Eur. Phys. J. C}
  {\bf 7} (Feb., 1999) 347--354.

\bibitem{Bergerhoff1998}
B.~Bergerhoff, {\it Critical behavior of $\phi^4$-theory from the thermal
  renormalization group},  {\em Phys. Lett. B} {\bf 437} (Oct., 1998) 381--389.

\bibitem{Aurenche1992}
P.~Aurenche and T.~Becherrawy, {\it A comparison of the real-time and the
  imaginary-time formalisms of finite-temperature field theory for 2, 3 and
  4-point green functions},  {\em Nucl. Phys. B} {\bf 379} (July, 1992)
  259--303.

\bibitem{Landsman1987}
N.~Landsman and C.~van Weert, {\it Real- and imaginary-time field theory at
  finite temperature and density},  {\em Phys. Rep.} {\bf 145} (Jan., 1987)
  141--249.

\bibitem{Ezawa1991}
H.~Ezawa, {\em Thermal Field Theories}.
\newblock North-Holland Delta Series. Elsevier Science, Burlington, 1991.
\newblock Description based upon print version of record.

\bibitem{Bellac2000}
M.~L. Bellac, {\em Thermal field theory}.
\newblock Cambridge monographs on mathematical physics. Cambridge University
  Press, Cambridge, 1. paperback ed. (with corrections)~ed., 2000.
\newblock Includes bibliographical references. - Originally published: 1996.

\bibitem{Mustafa2023}
M.~G. Mustafa, {\it An introduction to thermal field theory and some of its
  application},  {\em The European Physical Journal Special Topics} {\bf 232}
  (July, 2023) 1369--1457.

\bibitem{Matsumoto1984}
H.~Matsumoto, Y.~Nakano, and H.~Umezawa, {\it Renormalization group at finite
  temperature},  {\em Phys. Rev. D} {\bf 29} (mar, 1984) 1116--1124.

\bibitem{Fujimoto1986}
Y.~Fujimoto, K.~Ideura, Y.~Nakano, and H.~Yoneyama, {\it The finite temperature
  renormalization group equation in $\lambda \phi^4$ theory},  {\em Phys. Lett.
  B} {\bf 167} (Feb., 1986) 406--410.

\bibitem{Fujimoto1988}
Y.~Fujimoto and H.~Yamada, {\it Finite-temperature renormalization group
  equation in {QCD}. {II}},  {\em Phys. Lett. B} {\bf 200} (jan, 1988)
  167--170.

\bibitem{Baier1990}
R.~Baier, B.~Pire, and D.~Schiff, {\it High temperature behaviour of the {QCD}
  coupling constant},  {\em Phys. Lett. B} {\bf 238} (apr, 1990) 367--372.

\bibitem{Nakkagawa1987}
H.~Nakkagawa and A.~Ni{\'{e}}gawa, {\it Temperature dependence of the
  non-abelian gauge couplings at finite temperature},  {\em Phys. Lett. B} {\bf
  193} (jul, 1987) 263--267.

\bibitem{Nakkagawa1988}
H.~Nakkagawa, A.~Ni{\'{e}}gawa, and H.~Yokota, {\it Non-abelian gauge couplings
  at finite temperature in the general covariant gauge},  {\em Phys. Rev. D}
  {\bf 38} (oct, 1988) 2566--2578.

\bibitem{Braaten1990}
E.~Braaten and R.~D. Pisarski, {\it Resummation and gauge invariance of the
  gluon damping rate in hot {QCD}},  {\em Phys. Rev. Lett.} {\bf 64} (mar,
  1990) 1338--1341.

\bibitem{Frenkel1992}
J.~Frenkel, A.~V. Saa, and J.~C. Taylor, {\it Pressure in thermal scalar field
  theory to three-loop order},  {\em Phys. Rev. D} {\bf 46} (oct, 1992)
  3670--3673.

\bibitem{Arnold1994}
P.~Arnold and C.~Zhai, {\it Three-loop free energy for pure gauge {QCD}},  {\em
  Phys. Rev. D} {\bf 50} (dec, 1994) 7603--7623.

\bibitem{Parwani1995a}
R.~Parwani and H.~Singh, {\it Pressure of hotg2$\upvarphi$4theory at orderg5},
  {\em Phys. Rev. D} {\bf 51} (apr, 1995) 4518--4524.

\bibitem{Andersen2009}
J.~O. Andersen, L.~T. Kyllingstad, and L.~E. Leganger, {\it Pressure to
  orderg8loggof massless $\upphi$4theory at weak coupling},  {\em J. High.
  Energy Phys.} {\bf 2009} (aug, 2009) 066--066.

\bibitem{Andersen2000}
J.~O. Andersen, E.~Braaten, and M.~Strickland, {\it Massive basketball diagram
  for a thermal scalar field theory},  {\em Phys. Rev. D} {\bf 62} (jul, 2000)
  045004.

\bibitem{Haque2025}
N.~Haque and M.~G. Mustafa, {\it Hard thermal loop—theory and applications},
  {\em Progress in Particle and Nuclear Physics} {\bf 140} (Jan., 2025) 104136.

\bibitem{Romatschke2023}
P.~Romatschke, {\it Negative coupling $\phi^4$ on the lattice},  {\em arXiv}
  (2023).

\bibitem{Romatschke2024a}
P.~Romatschke, {\it Life at the landau pole},  {\em AppliedMath} {\bf 4} (Jan.,
  2024) 55--69.

\bibitem{Romatschke2024b}
P.~Romatschke, {\it Alternative to perturbative renormalization in ( 3+1
  )-dimensional field theories},  {\em Phys. Rev. D} {\bf 109} (June, 2024)
  116020.

\bibitem{Bandyopadhyay2016}
A.~Bandyopadhyay, C.~A. Islam, and M.~G. Mustafa, {\it Electromagnetic spectral
  properties and debye screening of a strongly magnetized hot medium},  {\em
  Phys. Rev. D} {\bf 94} (Dec., 2016) 114034.

\bibitem{Bandyopadhyay2016a}
A.~Bandyopadhyay, N.~Haque, M.~G. Mustafa, and M.~Strickland, {\it Dilepton
  rate and quark number susceptibility with the gribov action},  {\em Phys.
  Rev. D} {\bf 93} (Mar., 2016) 065004.

\bibitem{Peshier1998}
A.~Peshier, B.~Kämpfer, O.~P. Pavlenko, and G.~Soff, {\it Thermodynamics of
  the phi4-theory in tadpole approximation},  {\em Euro. Phys Lett.} {\bf 43}
  (aug, 1998) 381--385.

\bibitem{Laine2016}
M.~Laine and A.~Vuorinen, {\em Basics of Thermal Field Theory: A Tutorial on
  Perturbative Computations}.
\newblock Springer International Publishing, 2016.

\bibitem{Bugrij1995}
A.~I. Bugrij and V.~N. Shadura, {\it Three-loop contributions to the free
  energy of $\lambda \varphi^4$ qft},  {\em arxiv} (1995)
  [\href{http://arxiv.org/abs/hep-th/9510232v2}{{\tt hep-th/9510232v2}}].

\bibitem{Andersen2001a}
J.~O. Andersen, E.~Braaten, and M.~Strickland, {\it Screened perturbation
  theory to three loops},  {\em Phys. Rev. D} {\bf 63} (apr, 2001) 105008.

\bibitem{Peshier1994}
A.~Peshier, B.~Kämpfer, O.~Pavlenko, and G.~Soff, {\it An effective model of
  the quark-gluon plasma with thermal parton masses},  {\em Phys. Lett. B} {\bf
  337} (oct, 1994) 235--239.

\bibitem{Peshier1996}
A.~Peshier, B.~Kämpfer, O.~P. Pavlenko, and G.~Soff, {\it Massive
  quasiparticle model of the {SU}(3) gluon plasma},  {\em Phys. Rev. D} {\bf
  54} (aug, 1996) 2399--2402.

\bibitem{Shuryak2005}
E.~Shuryak, {\it What rhic experiments and theory tell us about properties of
  quark–gluon plasma?},  {\em Nucl. Phys. A} {\bf 750} (Mar., 2005) 64--83.

\bibitem{Bannur2006}
V.~M. Bannur, {\it Strongly coupled quark gluon plasma (scqgp)},  {\em J. Phys.
  G} {\bf 32} (May, 2006) 993--1002.

\bibitem{Goloviznin1993}
V.~Goloviznin and H.~Satz, {\it The refractive properties of the gluon plasma
  in {SU}(2) gauge theory},  {\em Zeit. für Phys. C} {\bf 57} (dec, 1993)
  671--675.

\bibitem{Arjun2024}
K.~Arjun et~al., {\it A revisit to running coupling constant in qed and qcd in
  the context of thermal field theory},  {\em unpublished}.

\end{thebibliography}\endgroup
\appendix
	\section{Two point function evaluation}
	\subsection{Diagram 1}\label{diagram1}
	An essential equation for deriving the Feynman diagram in the finite temperature imaginary time formalism (ITF) can be expressed as follows:
	\begin{equation}
		\begin{aligned}\label{importantone}
			T \sum_n \int \frac{\dd[D]p}{(2\pi)^D} \frac{1}{p^2+m^2+(2\pi n T)^2}  
			&=\int \frac{\dd[D+1]{P}}{(2\pi)^{D+1}} \frac{1}{P^2+m^2} + \int \frac{\dd[D]{p}}{(2\pi)^{D}} \frac{n_B(\beta \varepsilon_p)}{\varepsilon_p} \\
			&= \frac{(m^2)^{\frac{D-1}{2}}}{(4 \pi)^{\frac{D+1}{2}}} \Gamma \left( \frac{1-D}{2} \right) + S_{\frac{D-1}{2}}(m,T)
		\end{aligned}
	\end{equation}
	where $K(n,x)$ is the modified Bessel function of the second kind and $n_B(x)=(\exp(x)-1)^{-1}$,
	with $P^2=p_0^2+p^2$ and \(	S_N(m, T)=\left( \frac{1}{\pi} \right)\sum_{n=1}^\infty \left( \frac{m}{2\pi n \beta} \right)^N K_N(n m \beta) \).
	Using the above relation, it is possible to express
	\begin{equation}
		\begin{aligned}\label{Eq. Tadpole2}
			\left[ \lbrace -\lambda \rbrace \TheTadpole_\itf \right] &= -\lambda \int T \sum_{n= -\infty}^\infty \frac{1}{(2 \pi n T)^2+p^2+m^2} \DD{3}{p} \\
			&= -\lambda \int \frac{1}{P^2+m^2} \DD{4}{P} -\lambda S_1(m,T) \\
			&=\left[ \lbrace -\lambda \rbrace \TheTadpole_\qft \right] -\lambda S_1(m,T)
		\end{aligned}
	\end{equation}
	In the context provided, the braces $\lbrace \lambda \rbrace$ serve as labels rather than multiplication factors. Upon proceeding with dimensional regularization, the Feynman diagram undergoes a substitution $\lambda=g \mu^\epsilon$, resulting in
	\begin{equation}\label{Eq.Tadpole2-1}
		\left[\lbrace -g \mu^\epsilon \rbrace \TheTadpole_\itf \right] = \frac{m^2g}{(4 \pi)^2}\left[ \frac{2}{\epsilon}+ \clubsuit_2(m,\mu)   \right]+\mathcal{A}(\epsilon)+\mathcal{O}(\epsilon^2) -g \mu^{\epsilon} S_{1}(m,T)
	\end{equation}
	with 
	\begin{equation}
		\clubsuit_n(m,\mu) = \psi(n)+\ln \left( \frac{4 \pi \mu^2}{m^2} \right)
	\end{equation}
	and
	\begin{equation}
		\frac{			\mathcal{A}(\epsilon)}{\epsilon} = \frac{1}{4} \bigg[  {{[\clubsuit_2} (m,\mu)]^2} + \frac{\pi^2}{3} - \psi'(2) \bigg]
	\end{equation}
	, where we utilize a standard result from \cite{Kleinert2001}.
	The validity of \cref{Eq.Tadpole2-1} can be confirmed through a similar integral result found in \cite{Andersen2000,Peshier1998}. The primary distinction lies in our definition of 
	\begin{equation}
		\lambda \sumint f(p)=g \mu^{\epsilon} \sum_{n_p=-\infty}^\infty \int f(p) \frac{d^{N-\epsilon}p}{(2 \pi)^{N-\epsilon}}
	\end{equation}
	, whereas in \cite{Andersen2000,Peshier1998}, it is defined as
	\begin{equation}
		g^2 \sumint_{\bar{MS}} f(p)= g^2 \left( \frac{e^{\gamma }\mu^2}{4 \pi}  \right)^{\epsilon} \sum_{n_p=-\infty}^\infty \int f(p) \frac{d^{N-2\epsilon}p}{(2 \pi)^{N-2\epsilon}}
	\end{equation}
	\subsection{Diagram 2}\label{diagram2}
	\begin{equation}
		\begin{aligned}
			\lbrace \lambda^2 \rbrace \TheDoublebubble_\itf &=\int \lambda^2 T^2 \sum_{n_{p1}=- \infty}^\infty \sum_{n_{p2}=- \infty}^\infty
			\frac{1}{\omega_{n_{p1}}^2+\varepsilon_{p1}^2}  \left[\frac{1}{\omega_{n_{p2}}^2+\varepsilon_{p2}^2} \right]^2 \DD{3}{p_1} \DD{3}{p_2}  \\
			&= -\lambda T \int \sum_{n_{p1}=- \infty}^\infty \frac{1}{\omega_{n_{p1}}^2+\varepsilon_{p1}^2} \DD{3}{p_1} \times -\frac{\partial}{\partial m^2}
			\left[ -\lambda T \int \sum_{n_{p1}=- \infty}^\infty
			\frac{1}{\omega_{n_{p1}}^2+\varepsilon_{p1}^2} \DD{3}{p_1}  \right] \\
			&= \TheTadpole_\itf \times -\pdv{m^2} \TheTadpole_\itf \\
			&= \left[ \TheTadpole_\qft - \lambda S_1(m,T) \right] \times -\pdv{m^2} \left[ \TheTadpole_\qft-\lambda S_1(m,T) \right] \text{	( where we used  \cref{Eq. Tadpole2})} 
		\end{aligned}
	\end{equation}
	At $\lambda = g \mu^\epsilon$ and $\epsilon \to 0$, we can write
	\begin{align}
		\TheDoublebubble_\itf &=	\TheDoublebubble_\qft
		+\frac{g^2}{4 \pi}S_1(m,T)S_0(m,T) \\
		&-\frac{g}{4 \pi}S_0(m,T) \left[  \TheTadpole_\qft \right]
		+ g S_1(m,T) \frac{\partial}{\partial m^2}  \TheTadpole_\qft \nonumber
	\end{align}
	From \cite{Kleinert2001} we can express
	\begin{align}
		\TheDoublebubble_\qft &= -\frac{m^2g^2}{(4 \pi)^4} \left[ \frac{4}{\epsilon^2}+\frac{2 \left[ \clubsuit_1(m,\mu) + \clubsuit_2(m,\mu) \right]}{\epsilon} +\mathcal{O}(\epsilon^0) \right]\\ 
		&=-\frac{m^2g^2}{(4 \pi)^4} \left[ \frac{4}{\epsilon^2}+2 \frac{\psi(1)+\psi(2)}{\epsilon}-\frac{4}{\epsilon} \ln \left( \frac{m^2}{4 \pi \mu^2} \right)+\mathcal{O}(\epsilon^0) \right]
	\end{align}
	and from \cref{Eq.Tadpole2-1} and \cite{Kleinert2001}	
	\begin{equation}
		\frac{\partial}{\partial m^2} \left[ \TheTadpole_\qft \right] = \frac{g}{(4 \pi)^2}\left[ \frac{2}{\epsilon}+\clubsuit_1(m,\mu)  \right]+\mathcal{O}(\epsilon) = \frac{g}{(4 \pi)^2}\left[ \frac{2}{\epsilon}+\psi(1)+\ln \left( \frac{4 \pi \mu^2}{m^2}  \right) \right]+\mathcal{O}(\epsilon)
	\end{equation}
	\subsection{Diagram 3}\label{scatter_diagram}
	Utilizing Feynman diagrams corresponding to the four-point functions as outlined in \cite{Arjun2022}, we derived the renormalization constants and renormalization group function for the finite temperature imaginary time formalism (ITF). It has been demonstrated that the renormalization group function exhibits the same structure in both ITF and non-thermal quantum field theory (QFT). Although this diagram is not directly referenced in the main body of this paper, it holds significant importance in calculating the counterterm diagrams. In this context, we formulate using references \cite{Arjun2022,Kleinert2001}
	
	\begin{align}
		\lambda^2 T	\int \DD{3}{p} \  \sum_{n_p= -\infty}^\infty \frac{1}{\omega_{n_p}^2+\varepsilon_p^2} \frac{1}{\omega_{n_p-n_r}^2+\varepsilon_{p-r}^2} = \lambda^2 \int \frac{1}{P^2+m^2}\frac{1}{(P-R)^2+m^2} \DD{4}{P} \\
		+\lambda^2 \sum_{\sigma,\sigma_1=\pm 1} \int  \frac{n_B(\beta \varepsilon_p)}{2 \varepsilon_p \varepsilon_{p+r}} \frac{1}{\sigma_1 \varepsilon_p+\varepsilon_{p+r}+ i \sigma \omega_{n_r}} \DD{3}{p} \nonumber
	\end{align}
	Corresponding diagrammatic expression with $\lambda=g \mu^\epsilon$  at $\epsilon \to 0$ is
	\begin{align}
		\left[ \lbrace (g \mu^\epsilon)^2 \rbrace \ \TheScatter_{\itf} \right]= \left[ \lbrace (g \mu^\epsilon)^2 \rbrace \
		\TheScatter_{\qft,R_0=\omega_{n_r}} \right]+ (g \mu^\epsilon)^2 \ W(r,n_r)
	\end{align}
	with
	\begin{align}
		W(r,n_r)=\int \frac{d^3p}{(2 \pi)^3} \frac{2n_B(\beta \varepsilon_p)}{\varepsilon_p} \frac{r^2+ 2 pr \cos \theta+\omega_{n_r}^2}{(r^2+ 2 pr \cos \theta+\omega_{n_r}^2)^2+4 \varepsilon_p^2 \omega_{n_r}^2}
	\end{align}
	Utilizing dimensional regularization and drawing from the result in the standard textbook \cite{Kleinert2001}, we can express that 
	\begin{align}
		\left[ \lbrace g^2 \mu^\epsilon \rbrace \ \TheScatter_{\itf} \right]=& \frac{g^2\mu^\epsilon}{(4 \pi)^2} \left( \frac{2}{\epsilon}+\psi(1)+\int_0^1 dx \ln \left[ \frac{4 \pi \mu^2}{R^2x(1-x)+m^2} \right]|_{R_0=\omega_{n_r}}+\mathcal{O}(\epsilon) \right) \\ &+ g^2 \mu^{2 \epsilon} \ W(r,n_r)
	\end{align}
	with $R=[\omega_{n_r},r]$
	\subsection{Diagram 4}\label{A-4}
	The formulas provided in \cref{scatter_diagram} will prove beneficial as we divide the sunrise/sunset diagrams into subdiagrams. With references \cite{Arjun2022,Kleinert2001,Kapusta2006} for support, we can express the integral expression for the sunrise/sunset diagram in ITF as follows:
	
	\begin{equation}\label{sunset4}
		I=\TheSunrise_\itf=\lambda^2 T^2\int \sum_{n_{p}=-\infty}^\infty \sum_{n_{q}=-\infty}^\infty
		\frac{1}{\omega_{n_p}^2+\varepsilon_p^2} \frac{(2 \pi)^3}{\omega_{n_q}^2+\varepsilon_q^2}\frac{\delta^3(p+q+r+s)}{\omega_{n_p+n_q+n_s}^2+\varepsilon_{r}^2} \DD{3}{p} \DD{3}{q} \DD{3}{r}  
	\end{equation}
	
	From \cite{Bugrij1995,Andersen2000,Andersen2001a}, the integral can be expressed as $\text{I}=\text{I}_1+\text{I}_2+\text{I}_3$, where 
	\begin{align}
		\text{I}_1&=\TheSunrise_\qft
		=\int \frac{\lambda^2}{P^2+m^2} \frac{1}{Q^2+m^2} \frac{1}{R^2+m^2} (2 \pi)^4  \delta^4(P+Q+R+S) \DD{4}{P} \DD{4}{Q} \DD{4}{R} 
	\end{align}
	with $S=[\omega_{n_s},\vec{s}]$, and
	\begin{align}
		\text{I}_2=\int \DD{3}{p}\frac{3n_B(\beta \varepsilon_p)}{2 \varepsilon_p} \sum_{\sigma_1=\pm 1} \TheScatter\left( (P+S)^2 \right)_\qft \nonumber
	\end{align}
	with $P=[i \sigma_1 \varepsilon_p,\vec{p}], \ S=[\omega_{n_s},\vec{s}] $
	and
	\begin{align}
		\TheScatter_\qft(K^2)=\int \frac{\lambda^2}{R^2+m^2} \frac{1}{Q^2+m^2} (2 \pi)^4 \delta^4(R+Q+K) \DD{4}{R} \DD{4}{Q} 
	\end{align}
	However, the third term  $\text{I}_3$ can be written as
	\begin{align}
		\text{I}_3=\lambda^2 \int \DD{3}{p} \DD{3}{q}  \frac{3n_B(\beta \varepsilon_p) n_B(\beta \varepsilon_q)}{4 \varepsilon_p \varepsilon_q}  \times \sum_{\sigma_1, \sigma_2=\pm 1} \frac{1}{(i\sigma_1 \varepsilon_p+i\sigma_2 \varepsilon_q+\omega_{n_s})^2+(\vec{p}+\vec{q}+\vec{s})^2+m^2} \nonumber
	\end{align}
	with $	P=[i \sigma_1 \varepsilon_p, \vec{p}] \ Q=[i \sigma_2 \varepsilon_q, \vec{q}] , \ S=[\omega_{n_s},\vec{s}]$. Now by referring to the corresponding result from \cite{Kleinert2001} and \cref{scatter_diagram}, we can write
	\begin{align}
		\TheSunrise_\itf=\TheSunrise_{\qft,S_0=\omega_{n_s}} 
		+\int \frac{d^3p}{(2 \pi)^3} \frac{3n_B(\beta \varepsilon_p)}{2 \varepsilon_p} \sum_{\sigma_1} \TheScatter_{\qft}(P+S)^2
		+ \text{I}_3
	\end{align}
	where
	\begin{align}
		\TheSunrise_{\qft,S_0=\omega_{n_s}}=-g^2 \frac{m^2}{(4 \pi)^4} \left( \frac{6}{\epsilon^2} +\frac{S^2}{2 m^2 \epsilon} \right)
		-g^2 \frac{m^2}{(4 \pi)^4} \frac{6}{\epsilon} \left[ \frac{3}{2}+\psi(1)+ \ln \left( \frac{4 \pi \mu^2}{m^2} \right) \right]
		+\mathcal{O}(\epsilon)
	\end{align}
	with $S^2=\omega_{n_s}^2+s^2$.
	\begin{align}
		\lbrace g^2\mu^\epsilon \rbrace
		\sum_{\sigma_1}\TheScatter_\qft(P+S)^2 &=\frac{2g^2\mu^\epsilon}{(4 \pi)^2} \left( \frac{2}{\epsilon}+\psi(1)\right) \\
		&+  \sum_{\sigma = \pm 1} \frac{g^2 \mu^\epsilon}{(4 \pi)^2} \left( \int_0^1 dx \ln \left[ \frac{4 \pi \mu^2}{[(i \sigma \varepsilon_p+\omega_{n_s})^2+(p+s)^2]x(1-x)+m^2} \right] \right) 
		+\mathcal{O}(\epsilon) \nonumber
	\end{align}
	Let us simplify by setting \(S=0\), \begin{align}
		\lbrace g^2\mu^\epsilon \rbrace
		\sum_{\sigma_1} \TheScatter_{\qft}=\frac{2g^2\mu^\epsilon}{(4 \pi)^2} \left( \frac{2}{\epsilon}+\psi(1)\right)
		- \frac{2g^2 \mu^\epsilon}{(4 \pi)^2} \left( \int_0^1 dx \ln \left[ 1-x+x^2 \right] \right)
		+\frac{2g^2 \mu^\epsilon}{(4 \pi)^2}\ln \left( \frac{4 \pi \mu^2}{m^2} \right) 
		+\mathcal{O}(\epsilon)
	\end{align}
	Now, combining the results, we can express the terms that diverge (i.e., the pole terms) as follows:
	\begin{align}
		\mathcal{K} \left( \TheSunrise_\itf \right) &= \mathcal{K} \left( \TheSunrise_{\qft,k_0=\omega_{n_k}} \right)+3 S_{1}(m,T) \ \mathcal{K}  \left( \TheScatter_\qft \right) \\
		&= \mathcal{K} \left( \TheSunrise_{\qft,k_0=\omega_{n_k}} \right)+3gS_{1}(m,T) \ \pdv{m^2} \mathcal{K} \left(  \TheTadpole_\qft \right)  \nonumber
	\end{align}
	At the external momentum \(S=0\), and as \(\epsilon \to 0\), the integral result is
	\begin{equation}
		\begin{aligned}
			\TheSunrise_{\itf,S=0} &= \TheSunrise_{\qft,S=0}+3 S_{1}(m,T) \ \mathcal{K}  \left(\TheScatter_{\qft} \right) 
			\\
			&+ 3S_1(m,T) \frac{g^2 }{(4 \pi)^2} \left( \clubsuit_1(m,\mu) + 2-\frac{\sqrt{3} \pi}{3} \right)  +  \frac{3g^2m^2}{32 \pi^4}  Y(m,T) \\
			&= \TheSunrise_{\qft,S=0}+3 g S_{1}(m,T) \pdv{m^2} \mathcal{K}  \left(\TheTadpole_{\qft} \right) 
			\\
			&+ 3 S_1(m,T) \frac{g^2 }{(4 \pi)^2} \left( \clubsuit_1(m,\mu) + 2-\frac{\sqrt{3} \pi}{3} \right)  +  \frac{3g^2m^2}{32 \pi^4}  Y(m,T)
		\end{aligned}
	\end{equation}
	with 
	\begin{equation}
		Y(m,T) =	\int_0^\infty \int_0^\infty U(x) U(y) G(x,y) \ dx \ dy
	\end{equation}
	and
	\begin{align}
		&U(x)=\frac{\sinh(x)}{\exp \left( \beta m \cosh(x) \right)-1}, \ G(x,y)=\ln \left( \frac{1+2 \cosh(x-y)}{1+2 \cosh(x+y)} \frac{1-2 \cosh(x+y)}{1-2 \cosh(x-y)} \right), \\
		&\int_0^1 \ln(1-x+x^2) dx=\frac{\sqrt{3} \pi}{3}-2
	\end{align}
	The approximation can be verified using results from \cite{Andersen2001a}
	\section{Counterterms}\label{Bs}
	\subsection{Counterterm 1}\label{B-1}
	This counterterm $\countertermone$ does not directly appear in the main text; however, it holds essential significance for the calculation of four-point functions as well as certain two-point counterterm calculations \cite{Arjun2022}.
	From \cite{Kleinert2001}, the counter term for divergence for the four-point function derived is
	\begin{align}
		\countertermone_\qft = - \mu^\epsilon g c_g^1=-\frac{3}{2} \mathcal{K} \left[ \TheScatter_{\qft} \right] 
	\end{align}
	The corresponding diagram in imaginary time formalism \cite{Arjun2022} is
	\begin{align}
		\countertermone_\itf = - \mu^\epsilon g c_g^1=-\frac{3}{2} \mathcal{K} \left[ \TheScatter_{\itf} \right] 
	\end{align}
	From \cref{scatter_diagram}, it is evident that the diverging term is identical for the diagram in the imaginary time formalism and non-thermal QFT. Consequently, we can express
	\begin{equation}
		\countertermone_\itf = \countertermone_\qft = -\frac{3}{2} \mathcal{K}  \left[ \TheScatter_{\qft} \right] = -\mu^\epsilon g \frac{3g}{(4 \pi)^2}\frac{1}{\epsilon}
	\end{equation}
	The same pole term can be also expressed at $\epsilon \to 0$ as 
	\begin{align}
		\countertermone= - \frac{3}{2} g \mu^\epsilon \pdv{m^2} \mathcal{K} \left[ \lbrace -g \mu^\epsilon \rbrace \TheTadpole_\qft \right]
	\end{align}
	\subsection{Counterterm 2}\label{counterterm2}
	Following the same procedure and convention as outlined in \cite{Kleinert2001} , if we define the $*$ operator as the substitution of the counter term $-m^2c_{m^2}$ or $-\mu^\epsilon g c_g$, we can express the counter terms as:
	\begin{equation}
		\begin{aligned}
			\countertermtwo_\qft &= \countertermbtwo_\qft *  -\frac{1}{2} \mathcal{K} \left[ \TheTadpole_\qft \right] \\
			&=-g \mu^\epsilon \times - \pdv{m^2} \TheTadpole_{\qft} \times \frac{1}{2 g \mu^\epsilon} \mathcal{K} \left[\TheTadpole_{\qft}  \right]
		\end{aligned}
	\end{equation}
	From \cref{diagram1}, we have the relation 
	\begin{align}
		\mathcal{K} \left[  \TheTadpole_\itf \right]=\mathcal{K} \left[  \TheTadpole_\qft
		\right]
	\end{align}
	Therefore, for ITF, the corresponding derivation is 
	\begin{equation}
		\begin{aligned}
			\countertermtwo_\itf &= \countertermbtwo_\itf * -\frac{1}{2} \mathcal{K} \left[ \TheTadpole_\itf \right] = \countertermbtwo_\itf * -\frac{1}{2} \mathcal{K} \left[ \TheTadpole_\qft \right] \\
			&=- g \mu^\epsilon \frac{-\partial}{\partial m^2} \TheTadpole_\itf \times \frac{1}{2 g \mu^\epsilon} \mathcal{K} \left[ \TheTadpole_\qft \right] \\
			&=-g \mu^\epsilon \left(\frac{-\partial}{\partial m^2}{ \TheTadpole_\qft }-\frac{g S_0(m,T)}{4 \pi}\right) \times  \frac{1}{2g \mu^\epsilon} \mathcal{K} \left[ \TheTadpole_\qft \right] \\
			&=	\countertermtwo_\qft+\frac{g}{4 \pi} \frac{S_0(m,T)}{2} \mathcal{K}\left[  \TheTadpole_\qft \right]
		\end{aligned}
	\end{equation}
	So using the results from \cite{Kleinert2001}, the above equation can be reduced as
	\begin{equation}
		\begin{aligned}
			\countertermtwo_\itf = \frac{2m^2g^2}{(4 \pi)^4} \left[ \frac{1}{\epsilon^2}+\frac{\clubsuit_1(m,\mu)}{2 \epsilon}+\mathcal{O}(\epsilon^0) \right] +\frac{g^2 m^2 S_0(m,T)}{(4 \pi)^3} \frac{1}{\epsilon}
		\end{aligned}
	\end{equation}
	\subsection{Counterterm 3}\label{counter3}
	From \cite{Kleinert2001}, the calculation proceeds as
	\begin{equation}
		\begin{aligned}
			\countertermthree_\qft &= \TheTadpole_\qft \times \frac{-1}{g \mu^\epsilon}\times - \mu^\epsilon g c_g^1 \\
			&= \TheTadpole_\qft \times \frac{-1}{g \mu^\epsilon} \times -\frac{3}{2} \mathcal{K} \left[ \TheScatter_{\qft} \right]
		\end{aligned}
	\end{equation}
	the corresponding diagram made with results from Appendices \cref{Eq.Tadpole2-1,scatter_diagram}  is
	\begin{equation}
		\begin{aligned}
			\countertermthree_\itf &=  \TheTadpole_\itf \times \frac{-1}{g \mu^\epsilon} \times -\frac{3}{2} \mathcal{K} \left[ \TheScatter_{\itf} \right] \\
			&=\left[ \TheTadpole_\qft - g\mu^\epsilon S_1(m,T) \right] \times \frac{-1}{g \mu^\epsilon} \times -\frac{3}{2} \mathcal{K} \left[ \TheScatter_{\qft} \right] \\
			&=\countertermthree_\qft-\frac{3}{2}S_1(m,T) \mathcal{K} \left[ \TheScatter_{\qft} \right] \\
			&=\frac{6m^2g^2}{(4 \pi)^4} \left[ \frac{1}{\epsilon^2}+\frac{\clubsuit_2(m,\mu)}{2 \epsilon}+\mathcal{O}(\epsilon^0) \right] - \frac{3g^2 \mu^\epsilon}{(4 \pi)^2} \frac{1}{\epsilon} S_1(m,T)
		\end{aligned}
	\end{equation}
	The result can also be expressed as
	\begin{equation}
		\begin{aligned}
			\countertermthree_\itf = \countertermthree_\qft - \frac{3g \mu^\epsilon 	S_1(m,T)}{2} \pdv{m^2} \mathcal{K} \left[ \lbrace -g \mu^\epsilon \TheTadpole_\qft \rbrace \right]
		\end{aligned}
	\end{equation}
	\section{Vacuum Diagrams}
	\subsection{Vacuum diagram 1}\label{vacuumdiagram1}
	The analytic expression for the one-loop diagram in non-thermal QFT in $D$ dimension \cite{Kleinert2001} is
	\begin{equation}
		\begin{aligned}
			\TheVacuuma_\qft &= \int \DD{D}{P} \ln \left( P^2+m^2 \right) = \frac{2}{D} \left( \frac{m^2}{4 \pi} \right)^{D/2} \Gamma \left(1-D/2 \right) 
		\end{aligned}
	\end{equation}
	and that of ITF is
	\begin{equation}
		\TheVacuuma_\itf = T \sum_n \int \DD{D-1}{p} \ln \left( (2 \pi n T)^2+p^2+m^2 \right)
	\end{equation}
	Following the same procedures as in \cite{Kleinert2001} and doing some algebraic manipulations in \cref{importantone}
	\begin{equation}
		\TheVacuuma_\itf = \TheVacuuma_\qft - 4 \pi S_{D/2}(m,T)
	\end{equation}
	One might argue that the logarithm of the vacuum diagram displayed above lacks dimensional coherence, as its argument possesses dimensions of mass squared. Ideally, it should resemble a dimensionless form, such as $\ln \left( \varepsilon^2/\nu^2\right)$, where $\nu$ represents a mass scale. However, within the context of dimensional regularization, this manner of expressing the logarithm does not alter the integral. A detailed discussion, supported by the Veltman formula, can be found in \cite{Kleinert2001}. Since the integral is free of the vertex ($\lambda$), we will have to go through some other procedures described in \cite{Kleinert2001}, like multiplying the integral by a unit factor $\mu^\epsilon/\mu^\epsilon$ and expanding $\left( m^2/\mu^2\right)^{\epsilon/2}$ in powers of $\epsilon$ as $D \to 4-\epsilon$. Therefore
	\begin{equation}
		\begin{aligned}
			\frac{1}{2} \TheVacuuma_\qft = -\frac{1}{2} \frac{m^4}{\mu^\epsilon} \frac{1}{(4\pi)^2} \left[ \frac{1}{\epsilon} + \frac{\clubsuit_2(m,\mu)}{2}+  \frac{1}{4}   \right] + \mathcal{O}(\epsilon)
		\end{aligned}
	\end{equation}
	and
	\begin{equation}
		\frac{1}{2} \TheVacuuma_\itf = \frac{1}{2} \TheVacuuma_\qft - 2 \pi S_2(m,T)
	\end{equation}
	\subsection{Vacuum diagram 2}\label{vacuumdiagram2}
	The two-loop diagram corresponds to the Feynman integral for QFT, can be written as 
	\begin{equation}
		\begin{aligned}
			\TheVacuumb_\qft = -\lambda \left( \DD{4}{P} \frac{1}{P^2+m^2} \right)^2
		\end{aligned}
	\end{equation}
	The corresponding expression in ITF can be connected with that of QFT using \cref{Eq.Tadpole2-1} as
	\begin{equation}
		\begin{aligned}
			\TheVacuumb_\itf &= -\lambda \left( T \sum_{n} \int \frac{\dd[3]p}{(2\pi)^3} \frac{1}{\omega_n^2+\va{p}^2+m^2} \right)^2 \\ &=-\lambda  \left[ -\lambda^{-1} \times \lbrace -\lambda \rbrace \TheTadpole_\itf \right]^2 \\
			&=-\lambda \left( -\lambda^{-1} \left[ \lbrace -\lambda \rbrace \TheTadpole_\qft - \lambda S_1(m,T) \right] \right)^2 \\
			&=-\lambda \left(  \left[ -\lambda^{-1} \times \lbrace -\lambda \rbrace \TheTadpole_\qft \right]  +  S_1(m,T) \right)^2 \\
			&=\TheVacuumb_\qft + 2 \left[\lbrace -\lambda \rbrace \TheTadpole_\qft \right] S_1(m,T) - \lambda S_1^2(m,T)
		\end{aligned}
	\end{equation}
	From \cite{Kleinert2001}, the QFT diagram can be expanded in powers of $\epsilon$ at $\lambda \to g \mu^\epsilon$ as
	\begin{equation}
		\begin{aligned}
			\frac{1}{8}  \TheVacuumb_\qft  = -\frac{m^4g}{2(4\pi)^4 \mu^\epsilon}  \left[ \frac{1}{\epsilon^2} + \frac{1}{\epsilon} \clubsuit_2(m, \mu) +  \frac{A(\epsilon)}{\epsilon} + \left(\frac{\clubsuit_2(m, \mu)}{2}\right)^2 \right] + \mathcal{O}(\epsilon)
		\end{aligned}
	\end{equation}
	\subsection{Vacuum counterterm}\label{vacuumcounterterm}
	The vacuum counterterm can be derived using the ideas in \cite{Kapusta2006} as follows:
	\begin{equation}
		\begin{aligned}
			\TheVacuumCounterterm_\itf &= -\frac{1}{2} \mathcal{K} \left[ \TheTadpole_\itf \right] \times \frac{\partial}{\partial m^2} \left[ \TheVacuuma_\itf \right] \\
			&= -\frac{1}{2} \mathcal{K} \left[ \TheTadpole_\itf \right] \times \left[-\lambda^{-1} \lbrace -\lambda \rbrace \TheTadpole_\itf \right] \\
			&= -\frac{1}{2} \mathcal{K} \left[ \TheTadpole_\qft \right] \times \left[-\lambda^{-1} \lbrace -\lambda \rbrace \TheTadpole_\qft+S_1(m,T) \right] \\
			&=\TheVacuumCounterterm_\qft-\frac{1}{2} \mathcal{K} \left[ \TheTadpole_\qft \right]S_1(m,T)
		\end{aligned}
	\end{equation}
	with
	\begin{equation}
		\begin{aligned}
			\frac{1}{2} \TheVacuumCounterterm_\qft = \frac{m^4 g}{2(4 \pi)^4 \mu^\epsilon} \left[\frac{2}{\epsilon^2} + \frac{\clubsuit_2(m,\mu)}{\epsilon} + \frac{A(\epsilon)}{\epsilon} \right]+\mathcal{O}(\epsilon)
		\end{aligned}
	\end{equation}
	and
	\begin{equation}
		\frac{1}{2} \TheVacuumCounterterm_\itf = \frac{1}{2} \TheVacuumCounterterm_\qft - \frac{1}{4} \mathcal{K} \left[\lbrace -g \mu^\epsilon \rbrace \TheTadpole_\qft \right]S_1(m,T)
	\end{equation}


\end{document}